\documentclass{article} 
\usepackage{iclr2026_conference,times}

\usepackage{amsmath,amsfonts,bm}

\def\eqref#1{equation~\ref{#1}}

\def\1{\bm{1}}

\DeclareMathAlphabet{\mathsfit}{\encodingdefault}{\sfdefault}{m}{sl}
\SetMathAlphabet{\mathsfit}{bold}{\encodingdefault}{\sfdefault}{bx}{n}

\usepackage{hyperref}
\usepackage{url}

\usepackage{microtype}
\usepackage{graphicx}
\graphicspath{{figures/}}
\usepackage{subcaption}
\usepackage{booktabs} 
\usepackage{multirow} 
\usepackage{xcolor} 
\usepackage{enumitem} 
\usepackage{tikz} 

\usepackage{amsmath}
\usepackage{amssymb}
\usepackage{mathtools}
\usepackage{amsthm}

\usepackage[capitalize,noabbrev]{cleveref}

\usepackage{pifont}
\newcommand{\cmark}{\ding{51}}
\newcommand{\xmark}{\ding{55}}

\theoremstyle{plain}

\theoremstyle{definition}

\theoremstyle{remark}

\title{\textsc{WebPII}: Benchmarking Visual PII Detection for Computer-Use Agents}

\author{Nathan Zhao \\
Stanford University \\
\texttt{nathanzh@stanford.edu} \\
}

\iclrfinalcopy
\begin{document}

\maketitle


\begin{abstract}
Computer use agents create new privacy risks: training data collected from real websites inevitably contains sensitive information, and cloud-hosted inference exposes user screenshots. Detecting personally identifiable information in web screenshots is critical for privacy-preserving deployment, but no public benchmark exists for this task. We introduce \textsc{WebPII}, a fine-grained synthetic benchmark of 44,865 annotated e-commerce UI images designed with three key properties: extended PII taxonomy including transaction-level identifiers that enable reidentification, anticipatory detection for partially-filled forms where users are actively entering data, and scalable generation through VLM-based UI reproduction. Experiments validate that these design choices improve layout-invariant detection across diverse interfaces and generalization to held-out page types. We train \textsc{WebRedact} to demonstrate practical utility, more than doubling text-extraction baseline accuracy (0.753 vs 0.357 mAP@50) at real-time CPU latency (20ms). We release the dataset and trained model to support privacy-preserving computer use research.
\end{abstract}



\section{Introduction}

Computer use agents---language models that operate graphical user interfaces through vision and action---represent a significant capability advance toward general-purpose AI assistants. Unlike traditional web automation that operates on structured HTML or APIs, vision-based systems observe rendered web pages as images and produce mouse and keyboard actions to accomplish user goals. Recent systems such as Claude Computer Use~\cite{claudecomputeruse} and Gemini 2.5~\cite{gemini25} demonstrate purely vision-driven agents that can book flights, complete checkout flows, navigate e-commerce sites, and manage user accounts across arbitrary websites without access to DOM structure. As these systems scale from research prototypes to production deployments serving millions of sessions, their visual-first architecture creates fundamental privacy problems: every screenshot observation contains rendered PII, and standard cloud-hosted inference exposes sensitive user data during routine operation.

\begin{figure*}[t]
\centering
\begin{tikzpicture}
\node[anchor=north west,inner sep=0] (a) at (0,0) {\includegraphics[width=0.30\textwidth]{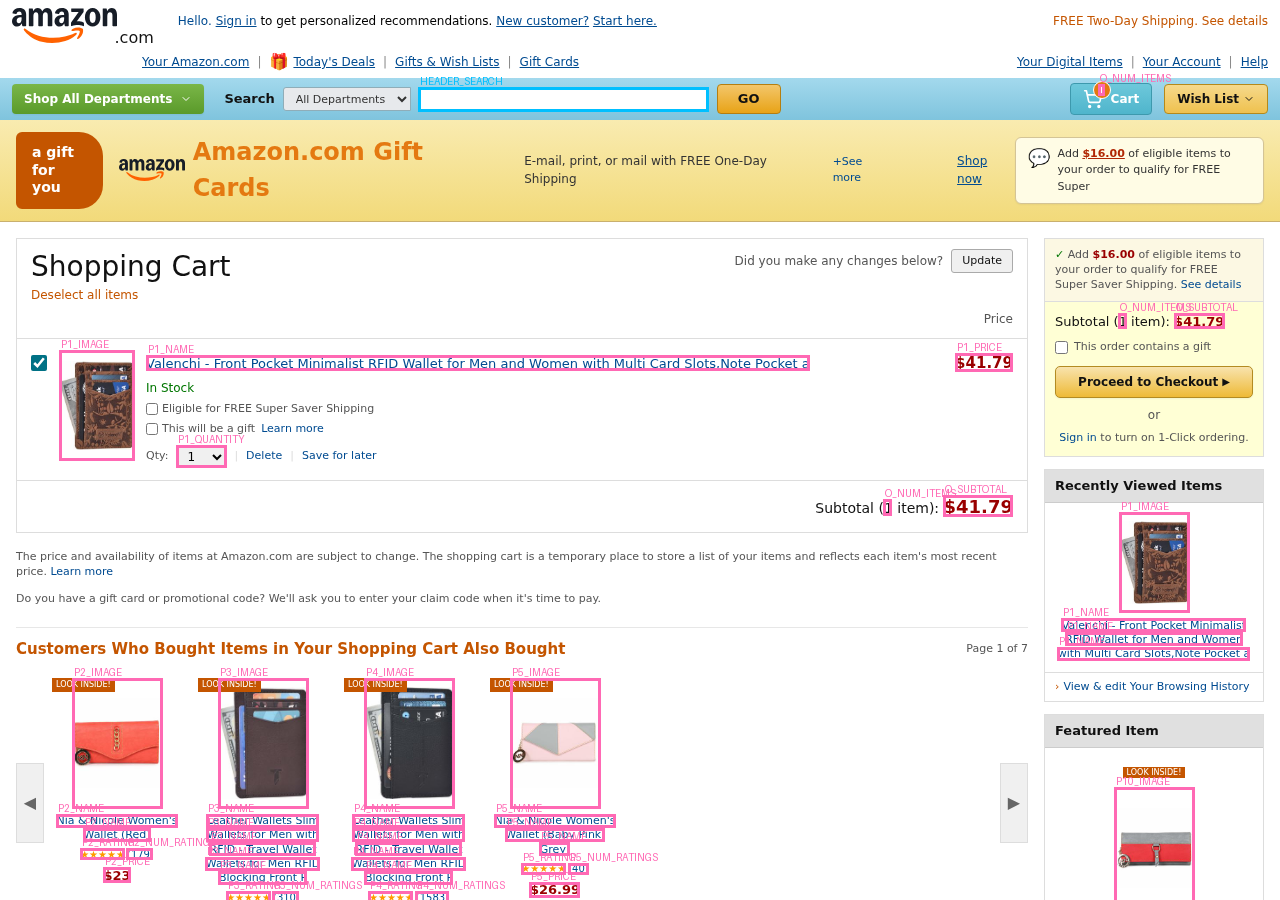}};
\node[fill=white,font=\footnotesize\bfseries,inner sep=1.5pt,anchor=north west] at ([shift={(2pt,-2pt)}]a.north west) {(a)};
\node[anchor=north west,inner sep=0] (b) at (a.north east) {\includegraphics[width=0.30\textwidth]{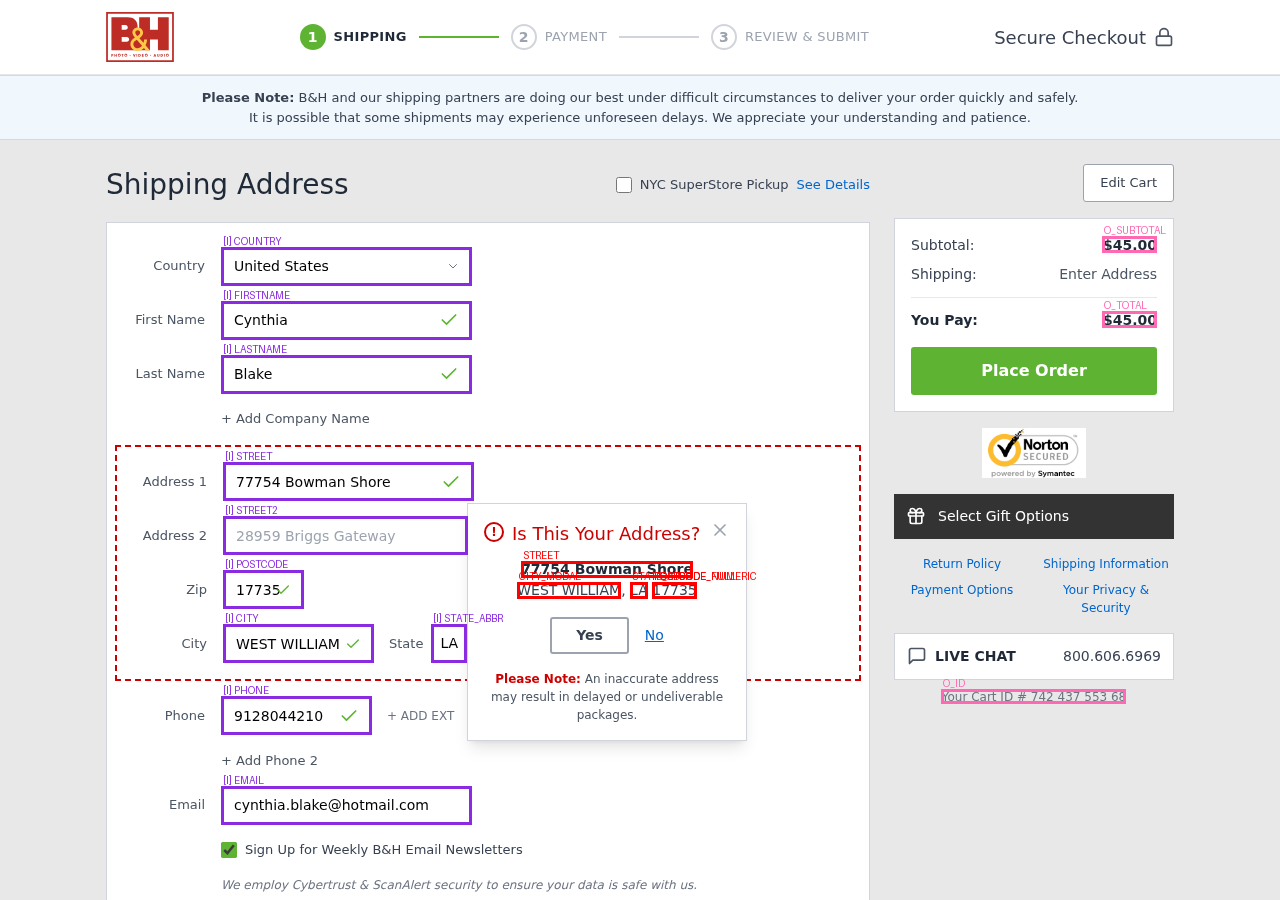}};
\node[fill=white,font=\footnotesize\bfseries,inner sep=1.5pt,anchor=north west] at ([shift={(2pt,-2pt)}]b.north west) {(b)};
\node[anchor=north west,inner sep=0] (c) at (b.north east) {\includegraphics[width=0.30\textwidth]{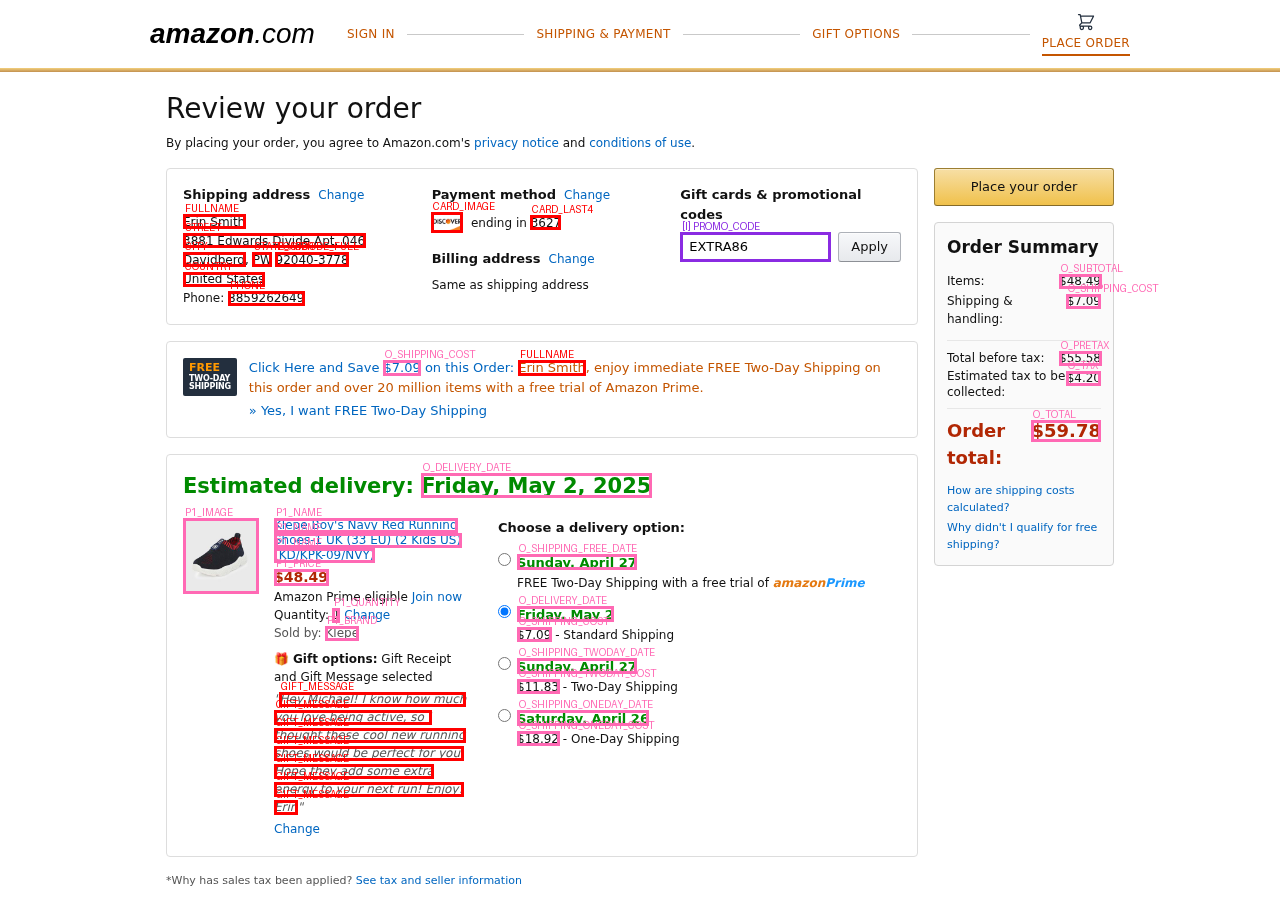}};
\node[fill=white,font=\footnotesize\bfseries,inner sep=1.5pt,anchor=north west] at ([shift={(2pt,-2pt)}]c.north west) {(c)};
\node[anchor=north west,inner sep=0] (d) at ([yshift={-2pt}]a.south west) {\includegraphics[width=0.30\textwidth]{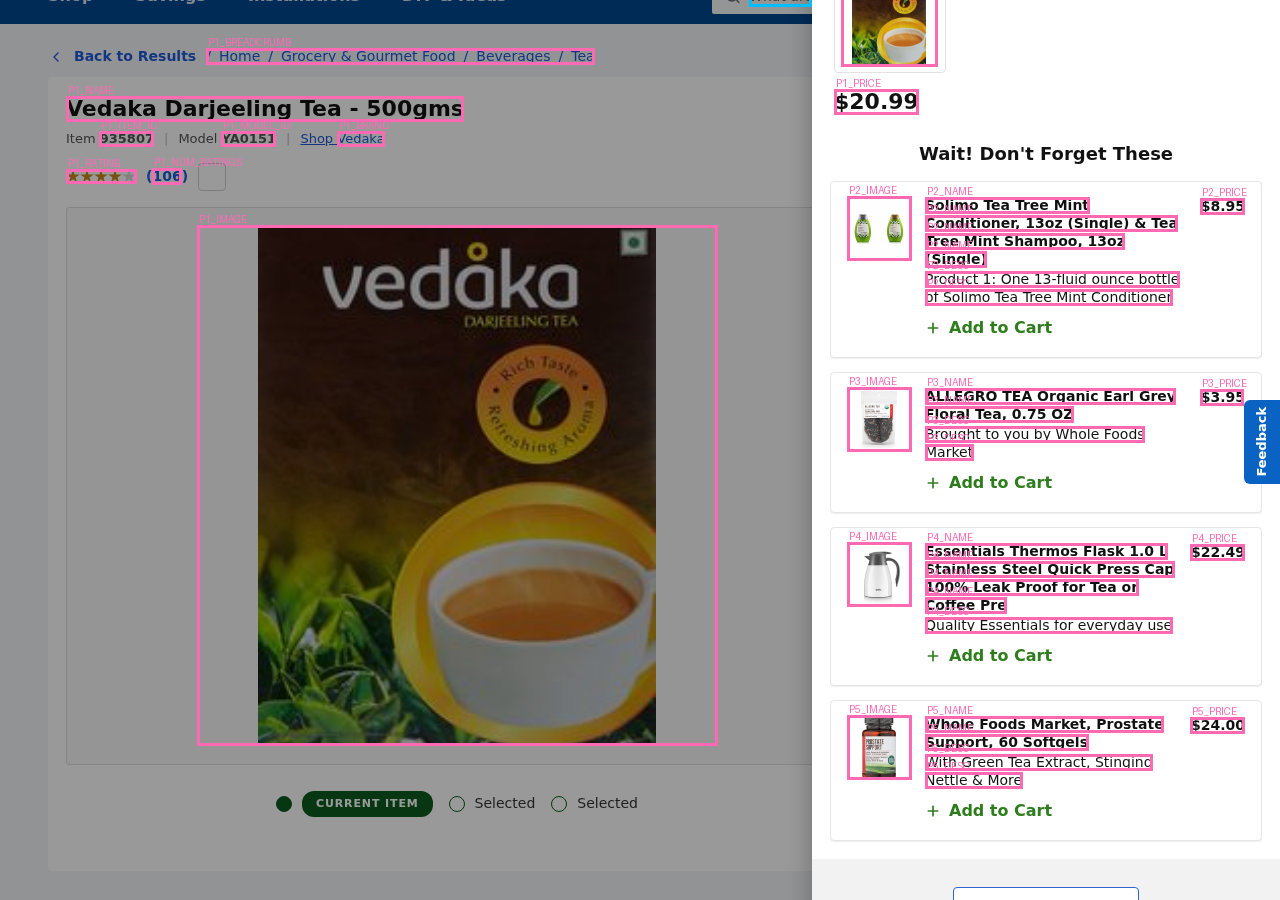}};
\node[fill=white,font=\footnotesize\bfseries,inner sep=1.5pt,anchor=north west] at ([shift={(2pt,-2pt)}]d.north west) {(d)};
\node[anchor=north west,inner sep=0] (e) at (d.north east) {\includegraphics[width=0.30\textwidth]{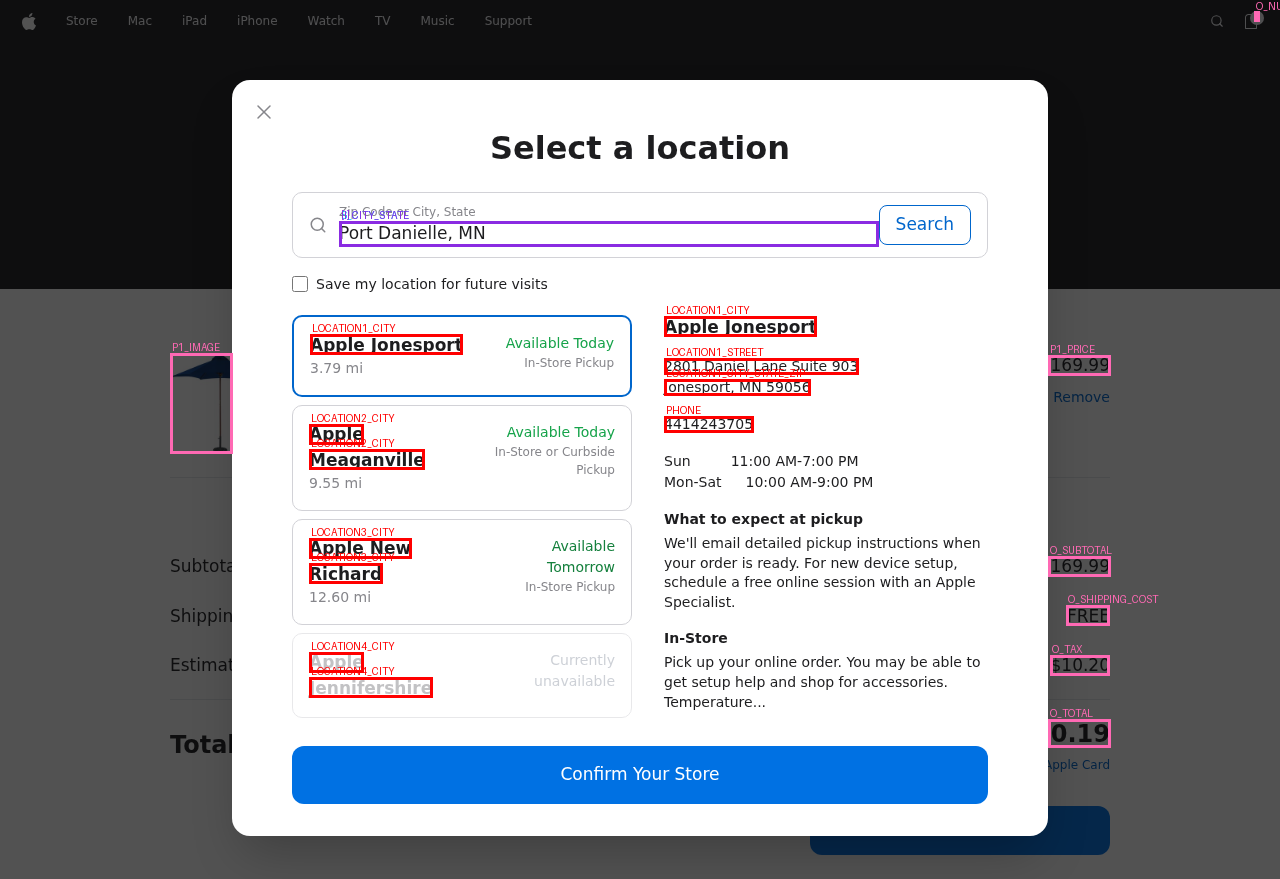}};
\node[fill=white,font=\footnotesize\bfseries,inner sep=1.5pt,anchor=north west] at ([shift={(2pt,-2pt)}]e.north west) {(e)};
\node[anchor=north west,inner sep=0] (f) at (e.north east) {\includegraphics[width=0.30\textwidth]{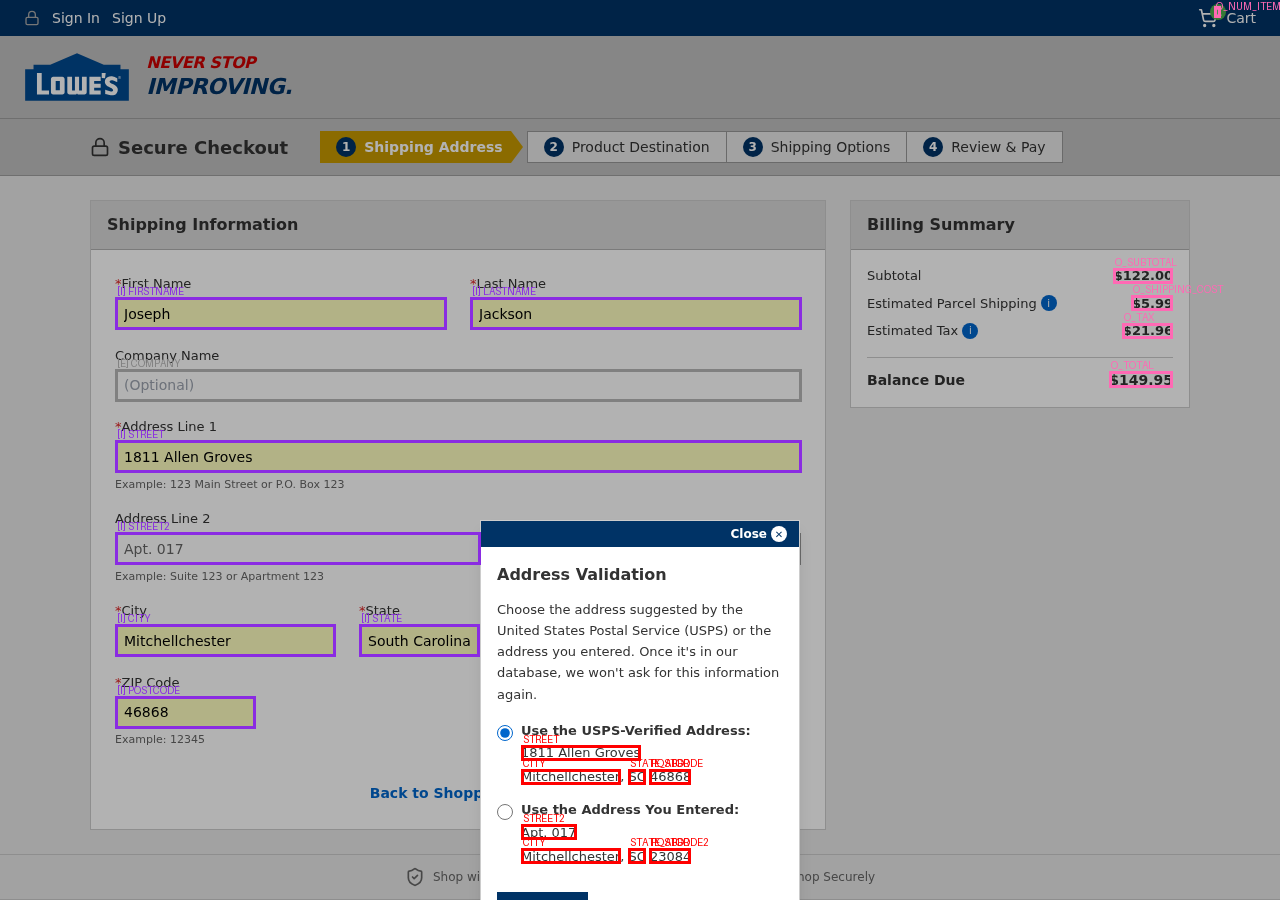}};
\node[fill=white,font=\footnotesize\bfseries,inner sep=1.5pt,anchor=north west] at ([shift={(2pt,-2pt)}]f.north west) {(f)};
\end{tikzpicture}
\caption{Sample images from \textsc{WebPII}, rendered with different injected data. The dataset captures the visual complexity of e-commerce interfaces: variable page heights reflecting diverse checkout flows and product displays (compare compact cart in (a) with extended layout in (d)), input fields and dropdown selectors, modal overlays with backdrops that occlude underlying content, ad hoc identifying information such as gift messages (c) and proposed pickup locations (e), and derived values requiring computation of taxes and totals. Bounding boxes respect occlusion boundaries. Pink indicates product annotations, purple denotes empty input fields, and red identifies PII.}
\label{fig:dataset_samples}
\end{figure*}
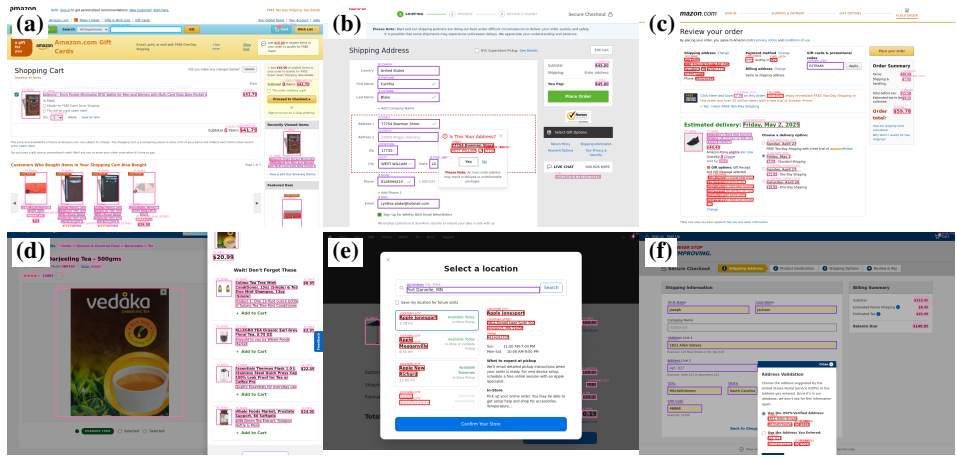

The privacy challenges span both training and inference. Training data collected from real websites inevitably contains PII that models memorize and leak~\cite{lukas2023pii, nasr2023extracting}, while cloud-hosted inference routinely exposes user screenshots. Existing mitigations are insufficient: sandboxed benchmarks~\cite{webarena, osworld} use fabricated data that does not transfer to real sessions, crowdsourced datasets~\cite{mind2web, weblinx} lack real-time authenticated content, agentic pipelines~\cite{opencua} lack visual PII detection, and federated approaches~\cite{mobilea3gent, fedmabench} still leak information through gradient updates.

Critically, \textbf{no public benchmark exists for visual PII detection in web interfaces}. Text-based PII systems~\cite{presidio, ai4privacy, panorama} operate on extracted strings, missing rendered content where sensitivity derives from visual context rather than surrounding words. Document-focused datasets~\cite{midv2020, docxpand} target fixed-layout identity documents with predictable field positions. Scene text datasets~\cite{cocotext, synthtext} localize text without distinguishing sensitive content from UI chrome.

\textbf{This paper presents a dataset-first contribution to address this gap.} We introduce \textsc{WebPII}, a fine-grained synthetic benchmark of e-commerce interface images designed with properties essential for visual PII detection in web interfaces. Our experiments validate key design choices---fill state diversity, data injection density, extended identifier taxonomy---demonstrating that the dataset enables layout-invariant PII detection across diverse e-commerce layouts.

Figure~\ref{fig:dataset_samples} illustrates the diversity of PII contexts across e-commerce layouts that we can dynamically vary through data injection. Beyond this visual diversity, the dataset introduces three validated properties:

\begin{enumerate}[leftmargin=*, itemsep=0.3em]
\item \textbf{Extended PII taxonomy.} Beyond traditional PII, \textsc{WebPII} annotates \textit{extended identifiers}---order numbers, tracking IDs, delivery dates, purchase histories---that enable re-identification but fall outside conventional PII definitions.

\item \textbf{Anticipatory detection.} Web interfaces expose PII as it is being typed. \textsc{WebPII} generates annotations at each progressive state of form-filling, enabling models to trigger redaction proactively rather than after sensitive data is fully visible.

\item \textbf{Scalable generation pipeline.} Our VLM-based approach dynamically generates diverse layouts and injects varied PII configurations, producing pixel-level bounding boxes without manual annotation.
\end{enumerate}

Our generation pipeline exploits VLM-based UI reproduction~\cite{claudecode, opencode} to generate functional frontend code from screenshots, embedding annotation attributes during code generation for programmatic bounding box extraction. Section~\ref{sec:study} documents the PII taxonomy underlying our annotations.

We validate the dataset's utility by training \textbf{\textsc{WebRedact}}, a visual detection model that more than doubles text-extraction baseline accuracy (0.753 vs.\ 0.357 mAP@50) at real-time CPU latency (20ms). Section~\ref{sec:experiments} presents ablation studies validating our design choices.

We release \textsc{WebPII}, \textsc{WebRedact}, and \textsc{WebRedact-Large}  to support development of privacy-preserving computer use systems.\footnote{Dataset and model available at: \url{https://webpii.github.io/}}


\section{The \textsc{WebPII} Dataset}
\label{sec:dataset}

E-commerce interfaces present PII challenges distinct from documents or scene text. While an email address in a scanned form appears as static pixels, the same email in a web UI may be rendered through JavaScript, styled with CSS, and wrapped in interactive elements. Moreover, web forms require \textit{anticipatory detection}---identifying sensitive fields before users finish typing, as privacy interventions should trigger during entry rather than after completion. Beyond traditional PII, these interfaces expose \textit{extended identifiers}---order IDs, tracking numbers, delivery dates---that enable reidentification despite not constituting traditional PII. Our benchmark must capture this complexity.

Generating annotated PII data at scale presents a dilemma: real screenshots contain real PII requiring manual annotation, while synthetic generation risks unrealistic layouts. We resolve this through \textit{reproduction with annotation injection}: collecting real e-commerce screenshots, then using vision-language models to recreate them as functional code with annotations embedded during generation. This produces pixel-accurate bounding boxes without manual labeling.

\subsection{Extended Identifiers in E-commerce}
\label{sec:study}

Beyond traditional PII (names, addresses, payment details), e-commerce interfaces display transaction-level attributes---order dates, merchants, item quantities, financial totals, delivery information---that enable reidentification despite not constituting traditional PII. Four credit card transactions containing just the merchant and date reidentify 90\% of 1.1M users~\cite{creditcardreidentification}; eight movie ratings with dates reidentify 99\% of Netflix users~\cite{netflixdeanon}. Cross-platform attacks link accounts by matching transaction patterns~\cite{Archie2018WhoS}.

These identifiers also reveal sensitive personal attributes---purchase patterns predict health conditions~\cite{Aiello2019LargescaleAH,Sasaya2026ValidatingBP}, socioeconomic status~\cite{Hashemian2017SocioeconomicCO}, and personal characteristics~\cite{Duhigg2012HowCL,Kosinski2013PrivateTA}---informing insurance underwriting~\cite{Allen2018Health}, data broker segmentation~\cite{Callanan2021TargetingVP}, and surveillance~\cite{Tokson2024Government,Kerr2021Buying,Sobel2023EndRunning}.

E-commerce interfaces also expose fields that are directly identifying yet absent from existing PII benchmarks: gift messages contain both sender and recipient names alongside personal notes; B2B purchase order numbers link individuals to corporate procurement systems and company names; delivery security codes enable package retrieval. Visual PII detection for agentic commerce must address these fields alongside traditional PII.

\newsavebox{\datainjtablebox}
\savebox{\datainjtablebox}{%
  \tiny
  \setlength{\tabcolsep}{1.5pt}
  \begin{tabular}{@{}lp{3.2cm}@{}}
  \toprule
  \textbf{Configuration Key} & \textbf{Value} \\
  \midrule
  \multicolumn{2}{@{}l}{\textit{Injected (Faker)}} \\
  \texttt{PII\_FULLNAME} & Marc Arnold \\
  \texttt{PII\_STREET} & 3400 Hester Green Suite 224 \\
  \texttt{ORDER\_DATE} & October 17, 2021 \\
  \texttt{ORDER\_DELIVERY\_DATE} & October 22, 2021 \\
  \midrule
  \multicolumn{2}{@{}l}{\textit{Injected (ABO)}} \\
  \texttt{PRODUCT1\_NAME} & 365 Everyday Value, Fragra... \\
  \texttt{PRODUCT1\_IMAGE} & \textit({product image}) \\
  \midrule
  \multicolumn{2}{@{}l}{\textit{LLM-Extracted}} \\
  \texttt{SHIPPING\_COST} & 5.99 \\
  \texttt{PRODUCT1\_BRAND} & 365 Everyday Value \\
  \midrule
  \multicolumn{2}{@{}l}{\textit{Randomized}} \\
  \texttt{PRODUCT1\_PRICE} & 4.49 \\
  \texttt{PRODUCT1\_RATING} & 4.7 \\
  \midrule
  \multicolumn{2}{@{}l}{\textit{Derived at Render Time}} \\
  \texttt{ORDER\_ID} & from \texttt{SEED} (647926) \\
  \texttt{ORDER\_TOTAL} & subtotal + ship + tax \\
  \bottomrule
  \end{tabular}
}

\begin{figure*}[t]
  \centering
  \begin{minipage}{0.48\textwidth}
    \includegraphics[width=\textwidth]{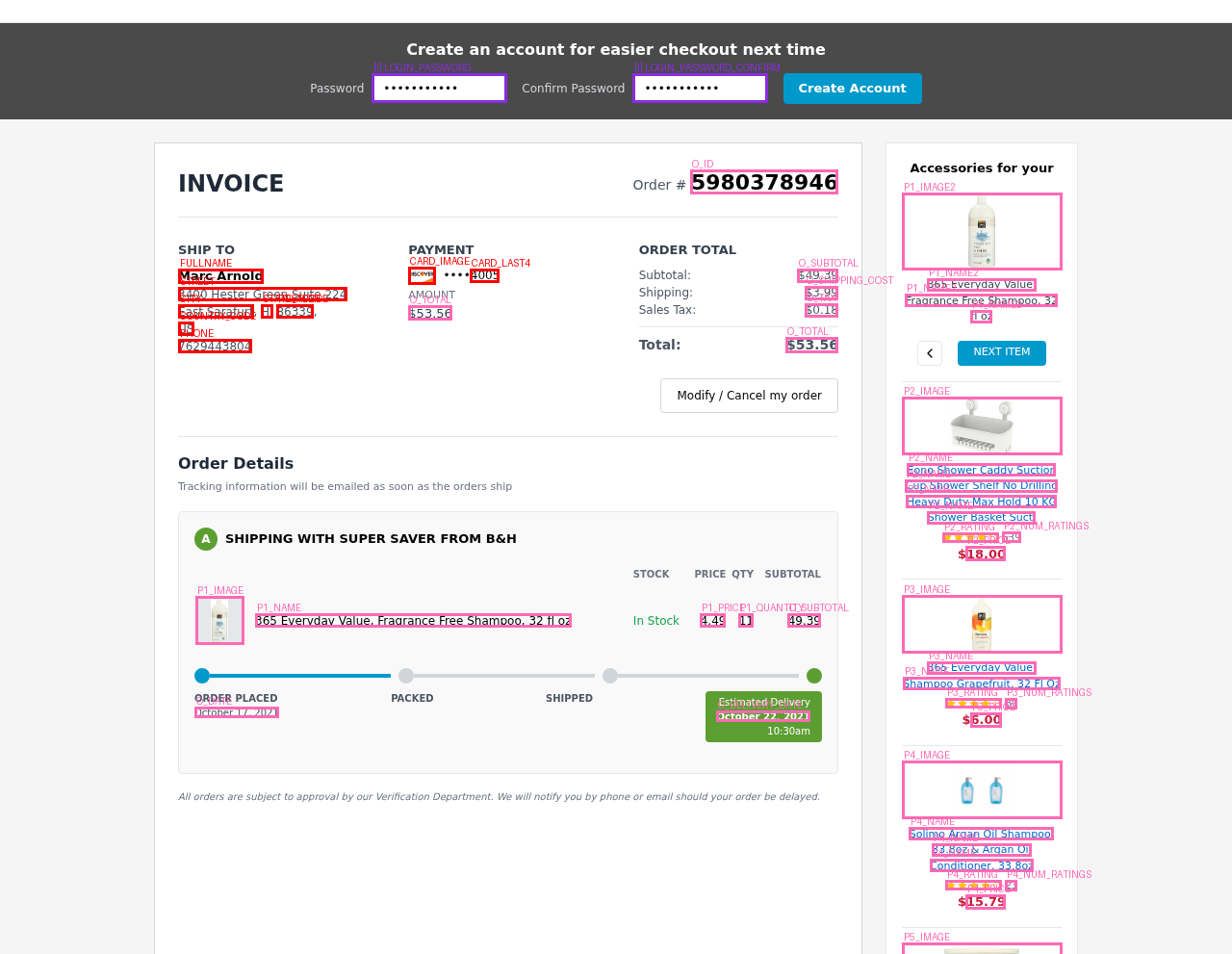}
  \end{minipage}%
  \hspace{0.02\textwidth}%
  \begin{minipage}{0.40\textwidth}
    \raggedright
    \resizebox{\linewidth}{!}{\usebox{\datainjtablebox}}
  \end{minipage}
  \vskip -0.1in
  \caption{Data injection maps configuration values to rendered UI elements. Left: annotated screenshot with bounding boxes. Right: selected subset of data injected for this page---Faker-generated PII, ABO product data, LLM-extracted metadata, and values derived at render time. The same layout rendered with different configurations produces diverse training examples with automatic annotations.}
  \label{fig:data_injection}
  \end{figure*}

\subsection{Data Collection and Generation}
\label{sec:pipeline}

We collected screenshots from 10 e-commerce brands across 19 page types---account dashboards, order history, order tracking, checkout flows, cart, billing address, delivery options, payment entry, gifting, store pickup, and product pages. Selection prioritized layouts exhibiting complex UI patterns: modal overlays, dynamic sidebars, embedded maps for store locators, cluttered multi-step forms, and responsive designs. We source multiple distinct layouts, including both current designs and prior designs that remain accessible, and so on, yielding 408 unique layouts. All of these original screenshots serve as visual targets for reproduction; we do not use them directly in the dataset.

For generation, we evaluated GPT-5.2~\cite{openai2025gpt52} with OpenCode~\cite{opencode} and Claude Opus 4.5~\cite{claudeopus45} with Claude Code~\cite{claudecode}; Claude exhibited superior rule-following on annotation conventions, requiring fewer iterations to reliably apply the schema.

The model generates React components, reproducing each screenshot with two enforced constraints: (1) all PII and product information must reference data variables rather than hardcoded values, and (2) every sensitive element must include a data attribute for annotation extraction. A Vite~\cite{vite} development server renders each component, and a Playwright~\cite{playwright} harness captures screenshots while extracting bounding boxes for all attributed elements via DOM queries. We provide the model with common e-commerce assets---company and payment method logos, security badges, shipping carrier icons, user avatar images, and map tiles---and pre-install icon libraries (Lucide, Heroicons, Phosphor) to avoid token-heavy inline SVG generation.\footnote{User avatars sourced from xsgames.co/randomusers; map tiles from openstreetmap.org (© OpenStreetMap contributors, ODbL).}

Generating accurate reproductions requires decomposing the task into specialized prompts, with context cleared between stages to prevent degradation. The pipeline proceeds in four stages: (1)~\textit{structure}, where the model generates a React component from the source screenshot, replacing all PII and product information with data variable references; (2)~\textit{attribute marking}, where a second pass tags each reference with annotation attributes for bounding box extraction; (3)~\textit{input handling}, where form fields are configured for partial-fill simulation (Section~\ref{sec:anticipatory}); and (4)~\textit{visual refinement}, where the rendered output is compared against the source in partitioned regions, with targeted fixes applied only to sections with discrepancies.

Beyond reproduction, the model extracts shipping costs and tax rates from source screenshots, derives order totals from item quantities, and infers platform-specific identifier formats. Human-in-the-loop validation corrects 39\% of layouts (viewport overflow, missing markup, initialization errors) via natural language instructions that delegate fixes back to Claude, typically requiring 1--2 iterations for any fixes. This validation amortizes across 25 data variants and all progressive fill states per layout. Section~\ref{app:annotation} provides more details on cost, human time spent handling, and decisions involved in creating the annotation pipeline.

\subsubsection{Data Injection}
\label{sec:label_categories}

The generated React components contain variable references rather than hardcoded values. At render time, data configurations populate these references with values from multiple sources: synthetic PII, product data, and attributes extracted or computed during generation. This enables the same layout to produce diverse screenshots with automatic annotations. We define three annotation types---\texttt{pii}, \texttt{product}, and \texttt{order}---corresponding to data attributes in our generation schema, illustrated in Figure~\ref{fig:data_injection}:

\textbf{PII annotations} (\texttt{data-pii}) cover traditional PII and context-specific fields. Values are generated via Faker~\cite{faker} with locale-appropriate formatting; context-specific fields (delivery instructions, security codes) use templated generation matching observed platform conventions.

\textbf{Product annotations} (\texttt{data-product}) capture displayed merchandise: names, descriptions, prices, images, ratings, and quantities. Product images are sourced from Amazon Berkeley Objects~\cite{abo} ({\raise.17ex\hbox{$\scriptstyle\sim$}}147K products, 400K images). Since ABO products oftentimes lack well-structured metadata, we clean the dataset for site-specific identifiers, placeholder images, and use GPT-4o-mini~\cite{openai2024gpt4omini} to extract brand names and item categories from titles and descriptions. For ``recommended'' and ``frequently bought together'' sections, we use fuzzy matching over product descriptions to retrieve semantically similar items. Values not provided by Faker or ABO are programmatically generated to follow realistic patterns, including review counts, ratings, and delivery dates.

\begin{figure}[t]
\centering
\begin{tikzpicture}
\node[anchor=north west,inner sep=0] (a) at (0,0) {\includegraphics[width=0.40\columnwidth,trim=0 200 0 200,clip]{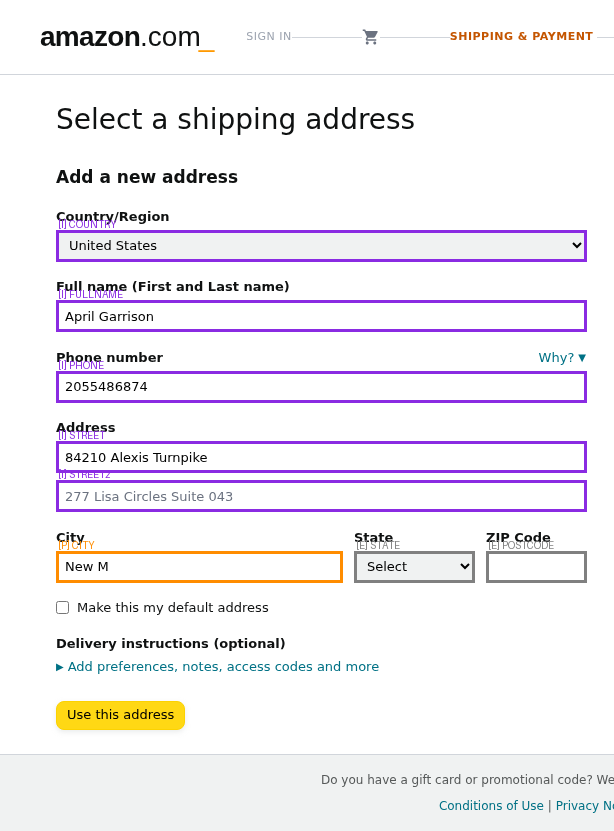}};
\node[fill=white,font=\footnotesize\bfseries,inner sep=1.5pt,anchor=north west] at ([shift={(2pt,-2pt)}]a.north west) {(a)};
\node[anchor=north west,inner sep=0] (b) at ([xshift=4pt]a.north east) {\includegraphics[width=0.40\columnwidth,trim=0 200 0 200,clip]{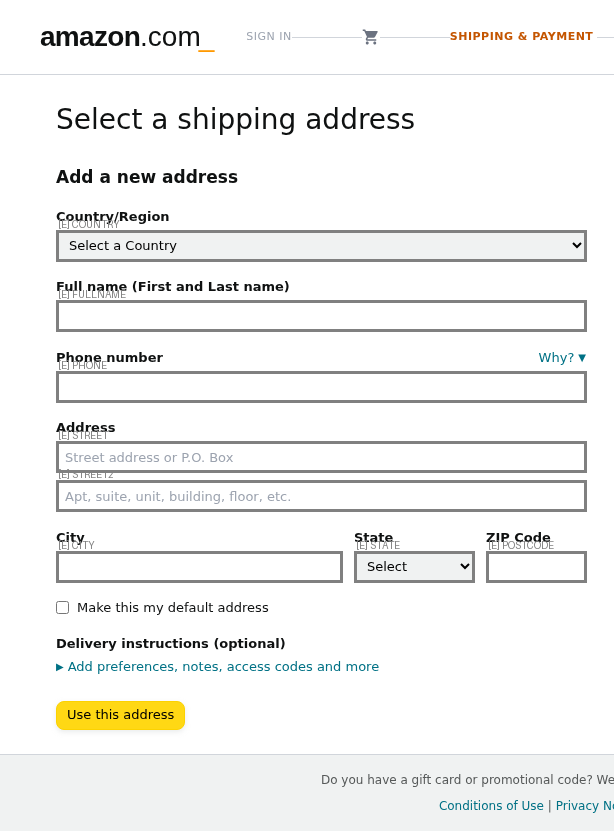}};
\node[fill=white,font=\footnotesize\bfseries,inner sep=1.5pt,anchor=north west] at ([shift={(2pt,-2pt)}]b.north west) {(b)};
\end{tikzpicture}
\vskip -0.05in
\caption{Form fill states for anticipatory detection. (a) Partial: mid-entry state with later fields incomplete (city field shows ``New M'' mid-typing). (b) Empty: pristine form with placeholder text and input field annotations. Yellow indicates partially filled fields; grey denotes empty fields.}
\label{fig:form_states}
\end{figure}

\textbf{Order annotations} (\texttt{data-order}) capture extended identifiers: order IDs, dates, tracking numbers, and financial totals. Order IDs and tracking numbers use seeded pseudo-random generation with platform-specific format templates. Shipping costs and tax percentages are extracted from the source screenshot during UI reconstruction, preserving realistic values for each layout. As shown in Figure~\ref{fig:data_injection}, derived values such as order subtotals and totals are computed at render time; the generation model must recognize these as derived and apply appropriate attributes.

Additional fields are generated via GPT-4o-mini when building data configuration files to produce realistic content: navigation breadcrumbs referencing browsing history (Figure~\ref{fig:dataset_samples}d), gift messages mentioning both recipient names and purchased items (Figure~\ref{fig:dataset_samples}c), and promotional copy. Each component $C_i$ is rendered with $m=25$ distinct data configurations, producing diverse screenshots with varied PII values, product images, and transaction details while maintaining automatic annotation alignment through DOM queries.

\subsection{Annotation Extraction}
\label{sec:annotation_attributes}

By requiring data attributes during generation, we extract pixel-coordinate bounding boxes directly from the rendered DOM using the browser's Range API to match text content with PII values. Wrapped text is split into separate bounding boxes per line via vertical overlap detection. This approach adapts to responsive layouts and ensures derived values receive proper attribution. Our extraction pipeline performs visibility analysis, excluding fully occluded elements and clipping partially visible ones to their visible extent.

\subsection{Anticipatory Detection}
\label{sec:anticipatory}

A key differentiator from text-based PII datasets is our support for \textit{anticipatory detection}---identifying PII before the user has finished typing. In deployed systems, privacy-preserving interventions should trigger as sensitive data is being entered, not after completion. For each component $C_i$ containing $N_i$ form fields, we generate anticipatory variants across progressive fill states: for stage $k$ (where $1 \leq k < N_i$), fields 1 through $k-1$ are fully filled, field $k$ is partially filled (mid-typing), and fields $k+1$ through $N_i$ remain empty. Figure~\ref{fig:form_states} illustrates two stages of this progression.

This systematic progression mirrors actual user interaction and ensures balanced representation across fill states. Models observe each field in three contexts: empty (before the user reaches it), partially filled (active input), and fully filled (after completion). Rather than biasing towards any particular fill state, the dataset provides equal exposure to all stages of form completion. Additionally, forms with optional fields are rendered with these fields randomly included or excluded across data injections, exposing models to structural variability in form layouts.

\subsection{Summary and Comparison}
\label{sec:statistics}

\textsc{WebPII} comprises 44,865 images spanning 10 e-commerce websites and 19 distinct page types, with 993,461 total bounding box annotations. This includes 10,200 base images with fully-filled forms, augmented by our anticipatory detection methodology that generates 28,653 partial-fill variants and 6,012 empty-form variants. Annotations span three categories (PII, product information, and order identifiers) across diverse HTML contexts (rendered text, input fields, and images). Detailed breakdowns of annotation density, category distribution, HTML element types, company distribution, and page type distribution appear in Appendix~\ref{app:statistics}.

\textsc{WebPII} is the first benchmark combining visual localization with semantic PII categories on rendered web interfaces, while annotating extended identifiers absent from prior work. Table~\ref{tab:dataset_comparison}  summarizes these distinctions.


\section{Experiments}
\label{sec:experiments}

\subsection{Experimental Setup}

\subsubsection{Data Splits}

The diversity of \textsc{WebPII} enables evaluation of generalization at different levels. We render each of the 408 unique layouts with 25 data injection variants (different PII values, addresses, and product information), and for layouts with input fields, generate the fill states described in Section~\ref{sec:anticipatory}: \emph{full}, \emph{partial}, and \emph{empty}. We evaluate three split strategies: \textbf{Test$_{\text{Cross-Page}}$} holds out 20\% of layouts randomly (82 layouts, 298 fill states), testing whether models learn layout-invariant features within a company's design system. \textbf{Test$_{\text{Cross-Company}}$} holds out all Amazon layouts (56 layouts, 152 fill states) while training on 11 other companies (352 layouts, 1,416 fill states), evaluating generalization to entirely new visual styles and brand identities. \textbf{Test$_{\text{Cross-Type}}$} holds out all receipt pages (20 layouts, 50 fill states) while training on 18 other page types (388 layouts, 1,518 fill states), measuring whether detection strategies transfer across functionally different page categories with distinct UI patterns. For all splits, we ensure no data leakage: the specific PII values, addresses, and product information in test images never appear in training.

\subsection{Baseline Methods}

\subsubsection{Text-Based Methods}

We evaluate text-based baselines using a two-stage pipeline: (1) OCR extraction to obtain text spans with bounding boxes, (2) classification to identify sensitive content. This approach mirrors existing PII detection systems that operate on extracted text rather than raw pixels. We compare three approaches: Presidio~\cite{presidio}, an NER-based system using pattern matching and named entity recognition; GPT-4o-mini for LLM-based classification; and document understanding models LayoutLMv3~\cite{layoutlmv3} and Donut~\cite{donut} that encode visual layout alongside text. All document understanding pipelines use GPT-4o-mini for final classification to ensure fair comparison. We evaluate multiple OCR engines Tesseract~\cite{smith2007tesseract}, EasyOCR~\cite{easyocr2024}, and PaddleOCR~\cite{cui2025paddleocr}, reporting the highest-accuracy configuration (LayoutLMv3) and fastest configuration (Tesseract + Presidio). Note that text-based approaches can only evaluate text content---product images are excluded as OCR provides no signal for visual elements. Appendix~\ref{app:ocr_llm_details} presents detailed ablations across all OCR engines, language models, and classification approaches.

\subsubsection{\textsc{WebRedact}}

We train object detection models on \textsc{WebPII} as \textsc{WebRedact}, targeting real-time inference suitable for live redaction. While \textsc{WebPII} contains fine-grained annotations for different PII types, we train on two classes grouped by visual appearance: \texttt{text} (all text-based elements including PII fields, product descriptions, and order information) and \texttt{image} (product images). This simplification improves detection reliability by providing more training examples per class and clearer visual distinctions between categories. Training uses 100 epochs, batch size 16, with standard augmentations (random crops, flips, color jittering). We train two model variants: \textsc{WebRedact} at 640$\times$640 resolution for real-time CPU inference, and \textsc{WebRedact-large} at 1280$\times$1280 resolution for higher accuracy when near-real-time constraints can be relaxed.

\begin{table}[t]
  \caption{Baseline results on Test$_{\text{Cross-Company}}$ (Amazon, full-fill images only). OCR+Presidio uses Tesseract; LayoutLMv3 uses GPT-4o-mini for classification.}
  \label{tab:main_results}
  \begin{center}
  \begin{small}
  \begin{sc}
  \begin{tabular}{lcc}
  \toprule
  Method & mAP@50 & Latency \\
  \midrule
  OCR + Presidio & 0.183 & 1.3s \\
  LayoutLMv3 + GPT-4o-mini & 0.357 & 2.9s \\
  \textsc{WebRedact} (ours) & 0.753 & \textbf{20ms} \\
  \textsc{WebRedact-large} (ours) & \textbf{0.842} & 312ms \\
  \bottomrule
  \end{tabular}
  \end{sc}
  \end{small}
  \end{center}
  \vskip -0.1in
  \end{table}

\subsection{Results}

Table~\ref{tab:main_results} presents main results on Test$_{\text{Cross-Company}}$, evaluating generalization to entirely new visual styles without seeing Amazon's design system during training. We evaluate text-based baselines only on full-filled images, as these approaches cannot identify empty or partially-filled input fields. Even on this favorable subset, both \textsc{WebRedact} variants substantially outperform text-based methods: \textsc{WebRedact} achieves 0.753 mAP@50, more than double the best text-based baseline (LayoutLMv3 at 0.357), while \textsc{WebRedact-large} reaches 0.842 mAP@50. Detailed failure mode analysis for OCR+LLM systems appears in Appendix~\ref{app:ocr_llm_details}.

\textsc{WebRedact} processes images in $\sim$20ms on mid-range consumer CPUs (Intel i5/AMD Ryzen 5), meeting real-time constraints for 30fps redaction, while \textsc{WebRedact-large} requires $\sim$312ms ($\sim$3fps). Both models use OpenVINO for CPU inference. Text-based methods are substantially slower: Tesseract~\cite{smith2007tesseract} OCR extraction alone (453ms, excluding classification) is slower than \textsc{WebRedact-large}'s full detection pipeline.

\subsection{Dataset Ablations}

We conduct ablations to validate \textsc{WebPII}'s design choices; full tables appear in Appendix~\ref{app:results}.

\textbf{Split strategies.} We compare three split strategies (Table~\ref{tab:split_comparison}). Test$_{\text{Cross-Page}}$ achieves 0.797 mAP@50, indicating models learn layout-invariant features within a company's design system. Test$_{\text{Cross-Company}}$ degrades to 0.753 when generalizing to Amazon's distinct visual style. Test$_{\text{Cross-Type}}$ shows the largest degradation (0.728), revealing that page-type-specific patterns transfer less effectively than company-specific design conventions.

\textbf{Fill state diversity.} Training on full screenshots alone achieves 0.771 mAP@50 (Table~\ref{tab:fill_ablation}). Adding empty screenshots \emph{without} partials degrades performance to 0.758, while full+partial achieves 0.797 (+0.026). Combining all three states achieves 0.825 mAP@50, demonstrating that partial fills provide essential intermediate visual grounding.

\textbf{Progressive fill density.} Increasing partial-fill stages per layout from 1 to 5 improves performance from 0.758 to 0.802 mAP@50 (Table~\ref{tab:progressive_fill_ablation}), with the strongest gains on partial-fill test images (0.774 to 0.835). Each additional stage exposes the model to new intermediate form states, providing non-redundant learning signal for field detection across the full spectrum of user interaction.

\textbf{Text variant density.} Increasing text variants per layout from 1 to 25 improves performance from 0.795 to 0.820 mAP@50 (Table~\ref{tab:variant_ablation_full}). Both precision and recall increase with more variants (0.805 to 0.842 and 0.713 to 0.731 respectively), confirming that diverse data injections---varying names, addresses, and products---improve generalization beyond layout diversity alone.


\section{Discussion}

\subsection{Why CPU-First Detection}

Our emphasis on lightweight CPU-executable models reflects deployment constraints for privacy-sensitive applications: on-device processing keeps screen content local rather than transmitting to cloud APIs~\cite{zhou2019edge}, continuous detection must not overload local compute, and high-volume scenarios (browser extensions, OS-level privacy layers) make per-inference GPU costs prohibitive. This aligns with broader trends toward on-device computer use agents~\cite{fara7b, uitars2, showui}, where privacy-preserving detection must operate within the same resource envelope.

\subsection{Limitations and Future Work}

Several limitations suggest directions for future work: domain scope is restricted to English-language e-commerce interfaces, with extension to other languages and domains (banking, healthcare, social media) requiring additional data collection; static image annotation does not capture challenges present in video streams with scrolling and state transitions; and our PII taxonomy conservatively labels all products and order identifiers as potentially identifying, which may cause over-redaction without contextual understanding.

Beyond privacy protection, \textsc{WebPII} enables applications in computer use agent systems. PII, product, and input region detection provides semantic grounding for agent inference---extending GUI grounding approaches~\cite{Lu2024OmniParserFP,Cheng2024SeeClickHG,Gou2024NavigatingTD,Yang2023SetofMarkPU,gelato2025,feizi2025grounding} to allow agents to condition behavior on field sensitivity. The synthetic generation pipeline's controllability could enable RL environment construction for agent training, where programmatic trajectory specification~\cite{Pahuja2025ExplorerSE,Wang2025AdaptingWA} enables scalable production of training data compared to expensive manual demonstration~\cite{mind2web,weblinx}.


\section{Conclusion}

We introduced \textsc{WebPII}, a fine-grained synthetic benchmark for visual PII detection in web interfaces, designed with three validated properties: extended PII taxonomy including transaction-level identifiers, anticipatory detection for partially-filled forms, and scalable generation through VLM-based UI reproduction. Our generation pipeline produces annotated data at scale by instrumenting LLM-driven UI reproduction with programmatic annotation extraction, requiring human validation for 39\% of layouts. Ablation studies validate these design choices, confirming that fill state diversity and data injection density each provide significant gains across split strategies.

We trained \textsc{WebRedact} to validate practical utility: the model more than doubles text-extraction baseline accuracy (0.753 vs 0.357 mAP@50) while achieving real-time CPU inference (20ms). As computer use agents transition from research prototypes to deployed systems, \textsc{WebPII} provides a foundation for integrating privacy protection into their design. We release the dataset and model to support privacy-preserving computer use research.

\section*{Impact Statement}

This work aims to improve privacy in computer use systems by enabling automated detection and redaction of sensitive information in web interfaces. While PII detection models could theoretically be misused for surveillance, actors with such intent can already extract PII through existing methods (foundation models, offline OCR) that do not require real-time visual localization. The availability of privacy-protective tools for the broader research community represents a net positive for user privacy.


\bibliography{references}
\bibliographystyle{iclr2026_conference}


\appendix


\newpage
\section{Dataset Statistics}
\label{app:statistics}

\begin{figure*}[!t]
\centering
\includegraphics[width=0.9\textwidth]{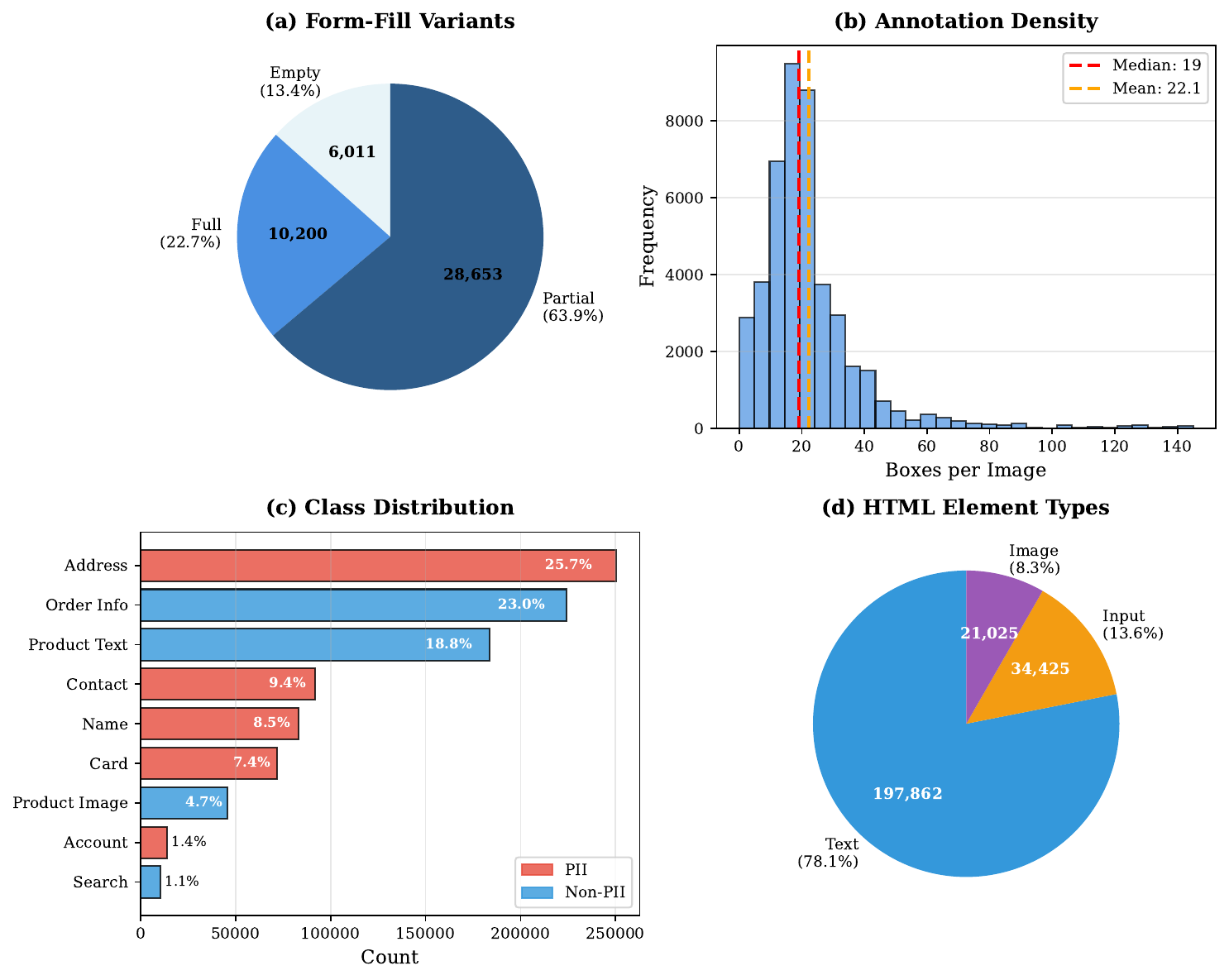}
\caption{Dataset composition and statistics. (a) Distribution across form-fill variants---empty forms (13.4\%), fully-filled forms (22.7\%), and partial-fill states (63.9\%)---enabling anticipatory detection training. (b) Annotation density distribution with median of 19 boxes per image (mean 22.1), ranging from 0 to 145 annotations per image. (c) Breakdown of all 9 annotation classes, with address (25.7\%), order info (23.0\%), and product text (18.8\%) dominating. PII classes (red) comprise 52.4\% of annotations, while non-PII classes (blue) comprise 47.6\%. (d) HTML element type distribution, showing most annotations target rendered text (78.1\%) versus input fields (13.6\%) and images (8.3\%).}
\label{fig:dataset_composition_appendix}
\end{figure*}

\begin{figure}[!t]
\centering
\includegraphics[width=0.75\textwidth]{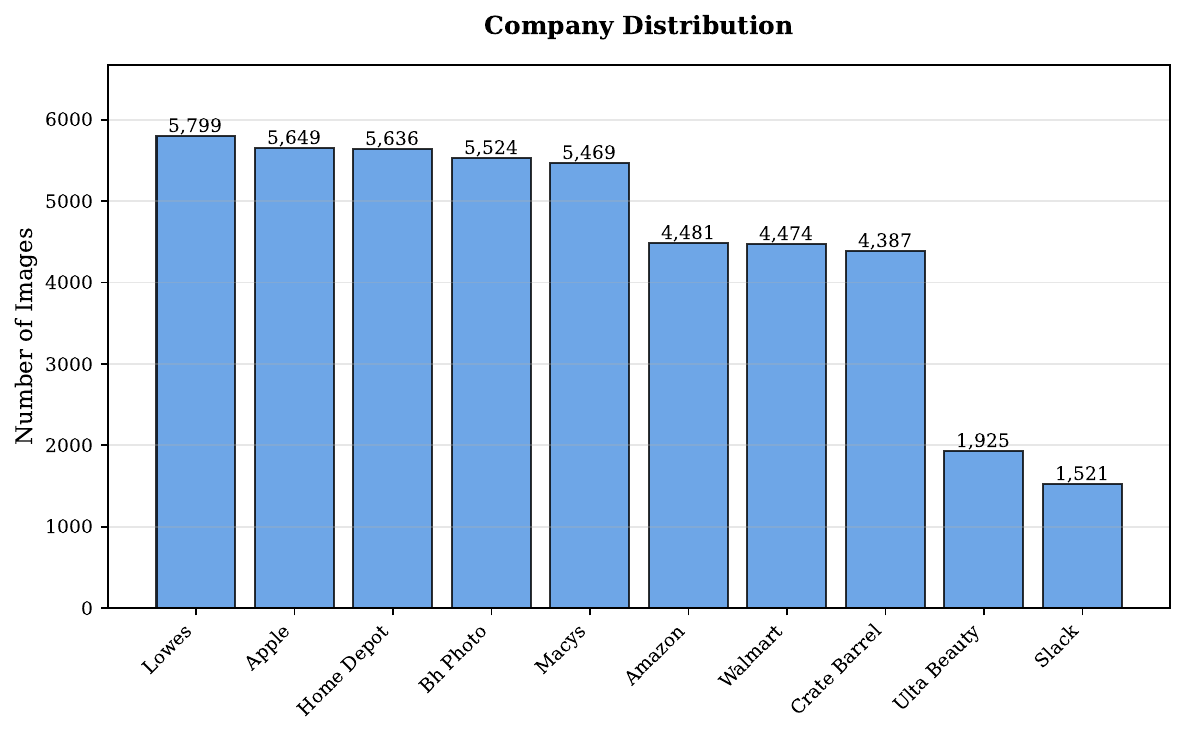}
\caption{Distribution of base images across 10 e-commerce companies. Apple and Amazon have the most coverage (1,400 images each), while Slack and Ulta Beauty represent smaller verticals (300 images each). Total base images: 10,200 (multiplied across variants to produce 44,865 total dataset images).}
\label{fig:company_distribution}
\end{figure}

\begin{figure}[!t]
\centering
\includegraphics[width=0.75\textwidth]{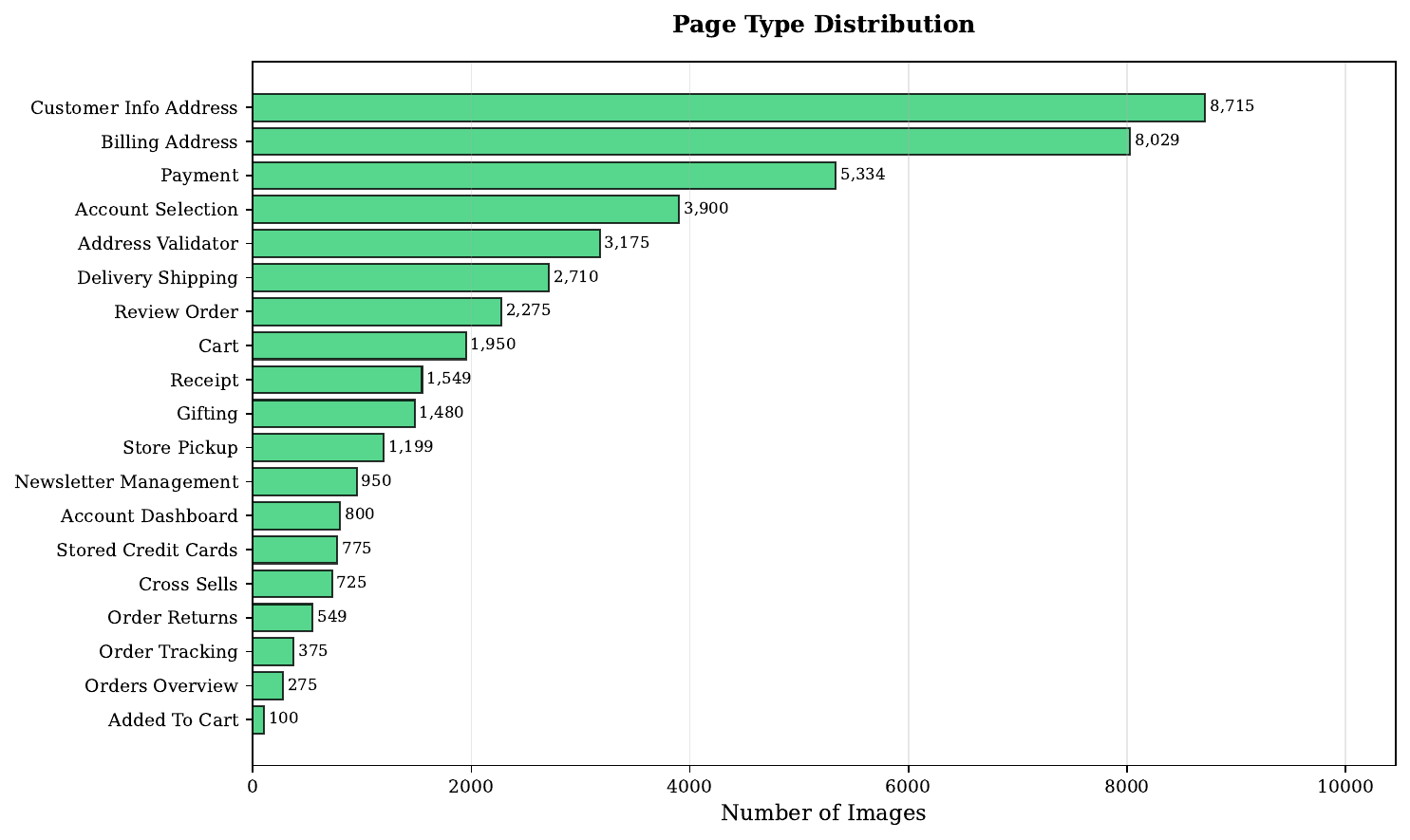}
\caption{Distribution of base images across 19 page types spanning checkout flows, account management, and product pages. Account selection (900) and delivery/shipping (875) pages dominate, representing critical moments where PII exposure is highest.}
\label{fig:page_type_distribution}
\end{figure}

The dataset comprises 44,865 images spanning 10 e-commerce websites and 19 page types. Figure~\ref{fig:dataset_composition_appendix} visualizes form-fill variants, annotation density, category breakdown, and HTML element types. Figures~\ref{fig:company_distribution}--\ref{fig:page_type_distribution} show company and page type distributions. Table~\ref{tab:dataset_comparison} compares \textsc{WebPII} against existing PII detection benchmarks.

The 52.4\%/47.6\% split between PII and non-PII annotations ensures models learn to distinguish sensitive user data from product and order metadata. Most annotations (78.1\%) target rendered text rather than input fields (13.6\%), reflecting that PII appears predominantly on confirmation and review pages where entered data is displayed, not just in the form fields where users type.

\begin{table}[!t]
  \caption{Comparison of PII detection benchmarks. \textsc{WebPII} is the first to combine visual localization, semantic categories, web interface targeting, extended identifiers, and anticipatory form state support.}
  \label{tab:dataset_comparison}
  \begin{center}
  \begin{scriptsize}
  \begin{sc}
  \begin{tabular}{lccccccc}
  \toprule
  Dataset & Domain & Size & Visual & Semantic & Web & Ext. & Antic. \\
  \midrule
  AI4Privacy~\cite{ai4privacy} & Text & 300K & \xmark & \cmark & \xmark & \xmark & \xmark \\
  PANORAMA~\cite{panorama} & Text & 385K & \xmark & \cmark & \xmark & \xmark & \xmark \\
  Nemotron-PII~\cite{nemotronpii} & Text & 100K & \xmark & \cmark & \xmark & \xmark & \xmark \\
  Gretel Finance~\cite{gretelfinance} & Text & 56K & \xmark & \cmark & \xmark & \xmark & \xmark \\
  BigCode PII~\cite{bigcode} & Code & 12K & \xmark & \cmark & \xmark & \xmark & \xmark \\
  PIILO~\cite{piilo} & Essays & 22K & \xmark & \cmark & \xmark & \xmark & \xmark \\
  MIDV-2020~\cite{midv2020} & ID Docs & 72K & \cmark & \cmark & \xmark & \xmark & \xmark \\
  DocXPand-25k~\cite{docxpand} & ID Docs & 25K & \cmark & \cmark & \xmark & \xmark & \xmark \\
  COCO-Text~\cite{cocotext} & Scene & 63K & \cmark & \xmark & \xmark & \xmark & \xmark \\
  SynthText~\cite{synthtext} & Scene & 800K & \cmark & \xmark & \xmark & \xmark & \xmark \\
  \midrule
  \textbf{\textsc{WebPII} (Ours)} & Web UI & 44K & \cmark & \cmark & \cmark & \cmark & \cmark \\
  \bottomrule
  \end{tabular}
  \end{sc}
  \end{scriptsize}
  \end{center}
  \vskip -0.1in
  \end{table}

\clearpage
\section{Human Quality Control Pipeline}
\label{app:annotation}

\subsection{Review Interface and Workflow}

Manually editing generated React code for hundreds of layouts would impose prohibitive overhead and require deep technical expertise. We instead developed a Flask-based web interface that accepts natural language correction instructions and delegates fixes to Claude. The interface operates on a queue system: layouts are displayed sequentially for review, with annotators able to examine the original screenshot, rendered reproduction, and all three annotated progressive fill states (empty, partial-fill, and fully-filled) to assess the full extent of data coverage. A code viewer provides access to the \texttt{App.jsx} source and \texttt{data.json} configuration for manual inspection when needed.

When an annotator identifies an issue, they submit a text instruction describing the required change. The system appends this instruction to a specialized fixing prompt that includes the current \texttt{App.jsx} source code, the original screenshot, and context from any previous fix iterations for that layout. Claude modifies the code and returns the corrected version. The interface then triggers re-rendering and re-annotation across all progressive fill states, allowing the annotator to verify the fix was applied correctly. Each layout underwent three review passes to ensure quality and provide redundancy in catching errors that might have been missed in earlier passes.

Refinement operates at the layout level rather than per-reproduction. Once a layout is corrected, the system re-screenshots across all three progressive fill states (empty, partial-fill, and fully-filled) to generate the final annotated dataset images, amortizing the cost of human review across multiple training samples.

We explicitly addressed the observed failure modes within the generation prompts themselves, providing specific instructions about viewport constraints, data attribute requirements, and form field initialization. However, rule-following remained imperfect for very nuanced cases---Claude would occasionally still produce viewport overflow or omit non-obvious data attributes despite explicit prompt guidance. Additionally, some errors were highly company-specific, and maintaining individually fine-tuned prompts per company would have undermined system scalability. We considered integrating more structured rubrics or checklists into the initial UI generation process to guide Claude through additional verification passes before human review. However, this would have introduced substantial costs through redundant forward passes that might not address the primary failure modes (viewport overflow detection, which requires human visual inspection, and context-dependent decisions about which non-obvious elements merit annotation). The conversational fixing interface proved more cost-effective by allowing targeted corrections only where needed rather than exhaustive verification on every layout.

\subsection{Refinement Statistics}

\begin{figure}[!ht]
\centering
\includegraphics[width=0.85\textwidth]{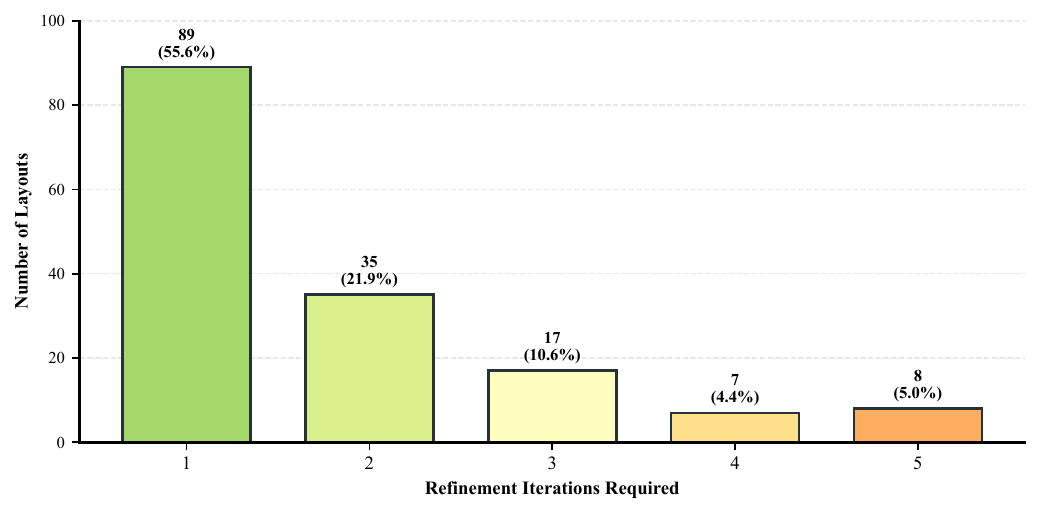}
\caption{Distribution of refinement iterations required per layout across 160 layouts requiring refinement. Over half (55.6\%) converged after a single correction iteration, while 21.9\% required two iterations. Four layouts required more than 5 iterations.}
\label{fig:refinement_iterations}
\end{figure}

The final dataset comprises 408 unique layouts. Of these, 160 (39.2\%) required human-initiated refinements, totaling 312 individual correction iterations. Figure~\ref{fig:refinement_iterations} shows the distribution of iterations required per layout. The majority (55.6\%) needed only a single correction, while 21.9\% required two iterations.

Total generation cost was \$649 using Claude Opus 4.5, comprising \$606 (93\%) for initial reproduction and \$43 (7\%) for refinements. This represents an average cost of \$1.50 per successfully generated layout including refinements, or \$1.41 per initial generation attempt. As expected, we observed notable variation according to page complexity.

Table~\ref{tab:fix_statistics} breaks down refinements by company and page type. Home Depot and Apple layouts required the most corrections (56 and 39 iterations respectively). Billing and payment pages accounted for 57 iterations across 29 layouts, the highest refinement density of any page type.

\begin{table}[h]
\caption{Distribution of human-initiated refinements across companies and page types, ordered by total fixes.}
\label{tab:fix_statistics}
\begin{center}
\begin{small}
\begin{sc}
\begin{tabular}{lcc}
\toprule
Category & Samples Refined & Total Fixes \\
\midrule
\multicolumn{3}{l}{\textit{By Company}} \\
Home Depot & 22 & 56 \\
Lowes & 23 & 44 \\
Amazon & 19 & 43 \\
Apple & 25 & 39 \\
Macy's & 13 & 39 \\
BH Photo & 21 & 33 \\
Walmart & 13 & 23 \\
Crate \& Barrel & 15 & 20 \\
Others & 9 & 15 \\
\midrule
\multicolumn{3}{l}{\textit{By Page Type}} \\
Billing/Payment & 29 & 57 \\
Cart & 23 & 38 \\
Customer Info/Address & 13 & 23 \\
Address Validator & 10 & 21 \\
Store Pickup & 8 & 20 \\
Added to Cart & 3 & 19 \\
Others & 74 & 134 \\
\midrule
\textbf{Total} & \textbf{160} & \textbf{312} \\
\bottomrule
\end{tabular}
\end{sc}
\end{small}
\end{center}
\vskip -0.1in
\end{table}

\subsection{Error Analysis and Failure Modes}

\begin{figure}[!ht]
\centering
\includegraphics[width=0.85\textwidth]{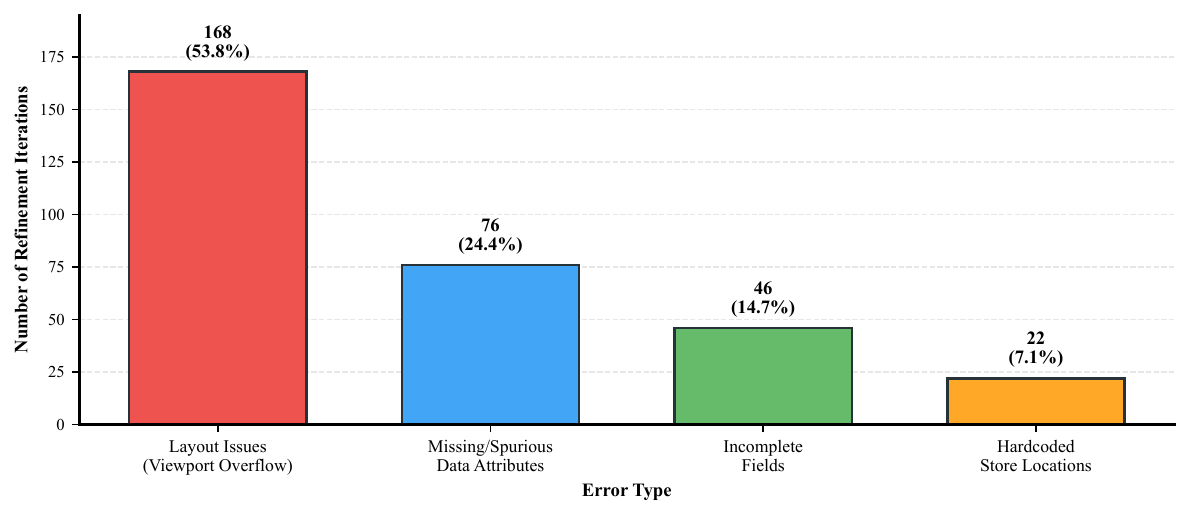}
\caption{Breakdown of 312 refinement iterations by error type across 160 layouts. Layout issues (53.8\%, primarily viewport overflow) dominated refinements. Data attribute errors (24.4\%) involved spurious attributes on decorative components (particularly in Apple's minimalist interfaces) and missing markup on product metadata fields. Incomplete fields (14.7\%) included dropdown initialization issues. Hardcoded store locations (7.1\%) required parametrization.}
\label{fig:fix_distribution}
\end{figure}

We manually categorized all 312 refinement iterations to understand failure modes (Figure~\ref{fig:fix_distribution}). Four primary error categories emerged:

\textbf{Layout issues (53.8\%).} The dominant failure mode involved UI elements being pushed outside the viewport width due to incorrect responsive sizing or fixed-width constraints. Claude's initial reproduction iterations included a visual refinement phase where a separate model invocation would compare rendered output to the original screenshot and suggest corrections. However, depsite iterating over various prompts, Claude remained lacking robust spatial reasoning in images---when elements overflowed horizontally beyond the viewport boundary, the model could not detect the issue from the rendered screenshot alone. These errors required human annotators to identify clipped content (e.g., right-aligned cart summaries cut off at 1400px width, sidebar filters extending beyond viewport) and explicitly instruct Claude to adjust container widths or implement proper responsive constraints.

\textbf{Missing or spurious data attributes (24.4\%).} These refinements predominantly involved product metadata fields like model numbers and item identifiers that were visually present but lacked proper \texttt{data-product} markup, likely do to the inapparent association with the general concept of PII. Apple interfaces exhibited a distinct failure mode: spurious \texttt{data-pii} or \texttt{data-product} attributes added to decorative elements or redundant container divs that should not have been annotated. These false positive annotations required explicit removal instructions.

\textbf{Incomplete fields (14.7\%).} Claude occasionally omitted input fields that did not match the common conceptual model of PII, requiring explicit addition instructions. A related issue involved selection dropdown defaults: Claude would sometimes hardcode dropdowns to specific values (e.g., ``United States'' for country selection) instead of initializing them to placeholder states (``Select a country''), which prevented proper partial-fill behavior where the dropdown should appear unselected. Note that certain fields were intentionally left unannotated---newsletter subscription email inputs, for example, were excluded from annotation as they represent marketing opt-ins rather than transaction-critical PII.

\textbf{Hardcoded store locations (7.1\%).} Some layouts contained store-specific location references that were not necessarily the user's personal location but could still be identifying or context-dependent---for example, ``Presque Isle's Lowes'' in headers where the store location was not the key focus of the page. Claude initially rendered these as literal placeholder text rather than parametrizing them, requiring explicit correction.

\subsection{Throughput and Pipeline Efficiency}

The Claude-mediated generation and refinement pipeline achieved substantial throughput despite the human review requirement. A single annotator processed approximately 100 layouts per 2 hours during the review phase. This high throughput was enabled by several factors. First, Claude handled the majority of actual code modifications---the annotator's role reduced to identifying errors and writing brief natural language correction instructions rather than performing manual code edits. Second, the interface provided immediate access to both the \texttt{App.jsx} source code and \texttt{data.json} configuration for manual inspection, allowing annotators to quickly diagnose issues by examining the underlying implementation. Third, the pipeline automated the full re-rendering and screenshot capture process---after submitting a fix, the system would rebuild the layout, capture screenshots across all three progressive fill states (empty, partial, full), run the annotation extraction, and present updated results within seconds. This fast iteration cycle eliminated manual overhead and enabled annotators to verify fixes immediately without context switching.

The primary bottleneck in dataset construction proved to be the collection of original e-commerce page screenshots rather than the generation or refinement process itself. Capturing diverse checkout flows, account dashboards, and payment pages across 10 major retailers required navigating authenticated sessions, filling realistic test data, and ensuring full-page screenshot capture across various UI states. Once source screenshots were collected, the automated pipeline consumed them rapidly. This architectural decision---delegating both generation and refinement to the model---proved essential for dataset feasibility, as manual HTML/CSS authoring or code-level debugging would have required orders of magnitude more human time.

\newpage
\section{\textsc{WebRedact-large} Ablations}
\label{app:small_model}

While \textsc{WebRedact} targets real-time inference at 30fps, \textsc{WebRedact-large} prioritizes accuracy while maintaining near-real-time performance ($\sim$3fps on CPU). This section provides complete analysis of \textsc{WebRedact-large} trained at 1280$\times$1280 resolution, including architectural comparisons, per-variant performance breakdowns, and inference latency measurements across hardware configurations.

\subsection{Architecture Comparison}

Table~\ref{tab:architecture_comparison} compares \textsc{WebRedact} and \textsc{WebRedact-large} on Test$_{\text{Cross-Page}}$ with 10 text variants.

\begin{table}[h]
\caption{Architecture comparison (2-class, cross-page, 10 text variants, trained on full+partial).}
\label{tab:architecture_comparison}
\begin{center}
\begin{small}
\begin{sc}
\begin{tabular}{lcccc}
\toprule
Model & Resolution & mAP@50 & Precision & Recall \\
\midrule
\textsc{WebRedact} & 640$\times$640 & 0.807 & 0.848 & 0.762 \\
\textsc{WebRedact-large} & 1280$\times$1280 & \textbf{0.909} & \textbf{0.868} & \textbf{0.834} \\
\bottomrule
\end{tabular}
\end{sc}
\end{small}
\end{center}
\vskip -0.1in
\end{table}

\textsc{WebRedact-large} achieves 0.909 mAP@50---a 12.6\% relative improvement over \textsc{WebRedact} (0.807). Both precision and recall improve substantially (+2.0pp and +7.2pp respectively), indicating that higher resolution enables more accurate field boundary detection and reduces false positives.

\subsection{Per-Variant Performance}

Table~\ref{tab:small_model_variants} shows per-variant performance breakdown for \textsc{WebRedact-large} across test fill states.

\begin{table}[h]
\caption{\textsc{WebRedact-large} per-variant performance (2-class, cross-page, 10 text variants).}
\label{tab:small_model_variants}
\begin{center}
\begin{small}
\begin{sc}
\begin{tabular}{lccc}
\toprule
Test Variant & mAP@50 & Precision & Recall \\
\midrule
full & 0.886 & 0.896 & 0.815 \\
partial & 0.943 & 0.897 & 0.878 \\
empty & 0.898 & 0.809 & 0.809 \\
\midrule
\textbf{Average} & \textbf{0.909} & 0.868 & 0.834 \\
\bottomrule
\end{tabular}
\end{sc}
\end{small}
\end{center}
\vskip -0.1in
\end{table}

The model achieves particularly strong performance on partial-fill images (0.943 mAP@50), where the progressive fill training provides rich learning signal. Performance on fully-filled forms (0.886) is slightly lower, likely due to increased visual complexity when all fields contain text.

\subsection{Inference Latency}

Table~\ref{tab:latency_comparison} compares inference latency between \textsc{WebRedact} and \textsc{WebRedact-large} on both CPU and GPU hardware.

\begin{table}[h]
\caption{Inference latency comparison (milliseconds per image).}
\label{tab:latency_comparison}
\begin{center}
\begin{small}
\begin{sc}
\begin{tabular}{lcc}
\toprule
Model & CPU (i5/Ryzen 5) & GPU (RTX 30xx) \\
\midrule
\textsc{WebRedact} & $\sim$20ms & $<$5ms \\
\textsc{WebRedact-large} & $\sim$312ms & $\sim$13ms \\
\bottomrule
\end{tabular}
\end{sc}
\end{small}
\end{center}
\vskip -0.1in
\end{table}

\textsc{WebRedact} satisfies real-time constraints on CPU ($<$33ms per frame for 30fps), while \textsc{WebRedact-large} requires 312ms ($\sim$3fps)---nearly 16$\times$ slower but still suitable for near-real-time applications where higher accuracy justifies relaxed frame rate requirements. On GPU hardware, \textsc{WebRedact-large} achieves 13ms latency, enabling real-time operation with GPU acceleration. For CPU-only deployment scenarios prioritizing maximum throughput, \textsc{WebRedact} remains the optimal choice.

\newpage
\section{Additional Results}
\label{app:results}

\subsection{Text-Based Baseline Details}
\label{app:ocr_llm_details}

\subsubsection{Document Understanding Models vs. OCR Pipelines}

We compare three categories of text-based baselines on Test$_{\text{Cross-Company}}$ (Amazon, full-fill images): document understanding models (LayoutLMv3~\cite{layoutlmv3}, Donut~\cite{donut}), OCR+LLM pipelines, and OCR+NER systems. Table~\ref{tab:doc_understanding_comparison} shows that document understanding models outperform simple OCR+LLM pipelines, with LayoutLMv3 achieving the highest mAP@50 (0.357) among all text-based approaches. To ensure fair comparison and bypass DocumentQA's single-span extraction limitation, all document understanding pipelines use GPT-4o-mini for final classification after visual/layout encoding rather than extractive question answering.

\begin{table}[h]
\caption{Document understanding models vs. OCR+LLM baselines on Test$_{\text{Cross-Company}}$ (Amazon, full-fill images). All methods use GPT-4o-mini for classification.}
\label{tab:doc_understanding_comparison}
\begin{center}
\begin{small}
\begin{sc}
\begin{tabular}{lccccc}
\toprule
Method & mAP@50 & Prec. & Rec. & F1 & Latency \\
\midrule
LayoutLMv3~\cite{layoutlmv3} & \textbf{0.357} & \textbf{58.5\%} & 45.1\% & \textbf{50.9\%} & 2.9s \\
Donut~\cite{donut} & 0.350 & 55.8\% & 43.6\% & 48.9\% & 3.5s \\
Tesseract~\cite{smith2007tesseract} + LLM & 0.302 & 49.8\% & \textbf{49.6\%} & 49.7\% & 2.8s \\
\bottomrule
\end{tabular}
\end{sc}
\end{small}
\end{center}
\vskip -0.1in
\end{table}

Document understanding models leverage visual layout information alongside text content, providing better localization on complex layouts like Amazon's dense product grids and multi-column interfaces. However, this advantage remains modest (+18\% mAP@50 over Tesseract~\cite{smith2007tesseract}+LLM), and the architectural mismatch between word-level detections and field-level ground truth limits all text-based approaches.

\subsubsection{OCR Engine Ablation}

Table~\ref{tab:ocr_engine_ablation} compares OCR engines with GPT-4o-mini classification. PaddleOCR~\cite{cui2025paddleocr} achieves the highest precision (78.7\%) but lowest recall (31.0\%), while Tesseract~\cite{smith2007tesseract} provides the best balance with competitive mAP@50 (0.302) at the fastest speed (2.8s). The narrow performance band (0.286--0.329 mAP@50) reveals that OCR quality is not the primary bottleneck---the architectural mismatch between word-level detections and field-level annotations limits all engines.

\begin{table}[h]
\caption{OCR engine ablation with GPT-4o-mini classification.}
\label{tab:ocr_engine_ablation}
\begin{center}
\begin{small}
\begin{sc}
\begin{tabular}{lccccc}
\toprule
OCR Engine & mAP@50 & Prec. & Rec. & F1 & Latency \\
\midrule
PaddleOCR~\cite{cui2025paddleocr} & \textbf{0.329} & \textbf{78.7\%} & 31.0\% & 44.5\% & 6.6s \\
Tesseract~\cite{smith2007tesseract} & 0.302 & 49.8\% & \textbf{48.6\%} & \textbf{49.2\%} & \textbf{2.8s} \\
EasyOCR~\cite{easyocr2024} & 0.286 & 67.6\% & 30.7\% & 42.2\% & 3.8s \\
\bottomrule
\end{tabular}
\end{sc}
\end{small}
\end{center}
\vskip -0.1in
\end{table}

The classification latency (2.8--6.6s) is dominated by LLM inference rather than OCR extraction. Table~\ref{tab:ocr_latency} shows OCR-only latencies. Tesseract~\cite{smith2007tesseract} completes in 453ms on CPU, while GPU-accelerated engines (EasyOCR~\cite{easyocr2024}, PaddleOCR~\cite{cui2025paddleocr}) are 1.6--4.7$\times$ slower despite hardware acceleration, indicating that OCR speed is not a primary bottleneck for text-based baselines, although still slower than our \textsc{WebRedact} and \textsc{WebRedact-large} models.

\begin{table}[h]
\caption{OCR engine latency (extraction only, no classification).}
\label{tab:ocr_latency}
\begin{center}
\begin{small}
\begin{sc}
\begin{tabular}{lc}
\toprule
OCR Engine & Latency \\
\midrule
Tesseract~\cite{smith2007tesseract} & \textbf{453ms} \\
EasyOCR~\cite{easyocr2024} & 715ms \\
PaddleOCR~\cite{cui2025paddleocr} & 2,143ms \\
\bottomrule
\end{tabular}
\end{sc}
\end{small}
\end{center}
\vskip -0.1in
\end{table}




\subsubsection{OCR + Presidio Baseline}

Table~\ref{tab:presidio_comparison} shows OCR+Presidio performance across engines. Presidio's rule-based NER achieves only 0.176--0.183 mAP@50, approximately 40\% worse than LLM-based classification. Tesseract~\cite{smith2007tesseract} provides the fastest configuration (1.3s), while PaddleOCR's~\cite{cui2025paddleocr} higher-quality extraction provides minimal benefit when the classifier lacks contextual understanding.

\begin{table}[h]
\caption{OCR + Presidio NER (no LLM).}
\label{tab:presidio_comparison}
\begin{center}
\begin{small}
\begin{sc}
\begin{tabular}{lccccc}
\toprule
OCR Engine & mAP@50 & Prec. & Rec. & F1 & Latency \\
\midrule
Tesseract~\cite{smith2007tesseract} & \textbf{0.183} & 41.7\% & \textbf{26.3\%} & \textbf{32.2\%} & \textbf{1.3s} \\
EasyOCR~\cite{easyocr2024} & 0.178 & 42.1\% & 28.3\% & 33.9\% & 2.3s \\
PaddleOCR~\cite{cui2025paddleocr} & 0.176 & \textbf{44.5\%} & 25.1\% & 32.1\% & 4.8s \\
\bottomrule
\end{tabular}
\end{sc}
\end{small}
\end{center}
\vskip -0.1in
\end{table}

\subsubsection{Document Question Answering}

An alternative to LLM classification is Document Question Answering (DocQA), where models answer explicit questions like ``What is the customer name?'' to extract PII. We evaluate two pure DocQA approaches---Donut-DocVQA and LayoutLM with extractive QA heads---against the hybrid configurations (LayoutLMv3 + LLM, Donut + LLM) described above. Table~\ref{tab:docqa_comparison} compares all approaches on Amazon test images.

\begin{table}[h]
\caption{Document QA vs. LLM classification on Amazon test images.}
\label{tab:docqa_comparison}
\begin{center}
\begin{small}
\begin{sc}
\begin{tabular}{lccccc}
\toprule
Method & mAP@50 & Prec. & Rec. & F1 & Latency \\
\midrule
LayoutLMv3~\cite{layoutlmv3} + LLM & \textbf{0.357} & \textbf{58.5\%} & 45.1\% & \textbf{50.9\%} & \textbf{2.9s} \\
Donut~\cite{donut} + LLM & 0.350 & 55.8\% & 43.6\% & 48.9\% & 3.5s \\
LayoutLM-QA & 0.139 & 24.1\% & \textbf{47.8\%} & 32.0\% & 1.5s \\
Donut-DocVQA & 0.117 & 30.8\% & 37.7\% & 33.9\% & 11.5s \\
\bottomrule
\end{tabular}
\end{sc}
\end{small}
\end{center}
\vskip -0.1in
\end{table}

DocQA approaches fail for several reasons. First, DocVQA models are trained to extract single answer spans per question (e.g., ``What is the date?'' $\rightarrow$ ``October 17''), not multiple items. This requires separate forward passes for each PII category (name, address, email, phone, card number, etc.), causing substantial latency: Donut's OCR-free architecture processes the full image for each question at $\sim$600ms per question (11.5s total), while LayoutLM's extractive QA operates on pre-extracted OCR text more efficiently (1.5s total). Second, both models are pre-trained on structured documents and forms, where PII appears in predictable locations with clear visual cues (labeled fields, tables). E-commerce screenshots present different visual hierarchies---gift recipient versus billing names, optional fields, promotional overlays---that deviate from the models' training distribution. Third, rigid question templates often mismatch actual page content, requiring flexible contextual understanding that the single-span extraction paradigm cannot provide. Fourth, LayoutLM-QA uses a randomly initialized QA head (not fine-tuned), further degrading span extraction quality.

LLM classification (sending all extracted text to GPT-4o-mini for contextual classification) achieves 2.6$\times$ higher mAP@50 (0.357 vs 0.139) than the best DocQA method while maintaining comparable or faster speed. The flexible classification paradigm handles diverse layouts without rigid question templates.

\subsubsection{Failure Modes}

The two-stage architecture creates cascading errors where OCR misdetections propagate to classification. On sparse pages with few ground truth elements, OCR misreads generic footer text or company names, causing the LLM to aggressively flag these misdetected strings as sensitive and producing false positives. Conversely, dense pages with complex product listings cause context rot in chunked LLM calls, where the model loses track of relevant fields as listings overwhelm the limited context window. Product names containing person names (e.g., ``Kelsey Villages'' as a street name versus ``Kelsey'' as a brand) are frequently misclassified, and gift messages containing potentially private data are sometimes missed.

\subsubsection{Classification Prompt}

The OCR + LLM baselines use GPT-4o-mini to classify extracted text spans. The classification prompt is:

\begin{quote}
\small
\texttt{You identify data values in e-commerce screenshot text.}

\texttt{Flag ANY text that is actual data (not a UI label):}\\
\texttt{Names, addresses, cities, states, zip codes}\\
\texttt{Emails, phone numbers, dates}\\
\texttt{Card numbers, CVV, expiry dates}\\
\texttt{Product names, brands, prices, quantities, ratings}\\
\texttt{Order totals, shipping costs, tracking numbers}\\
\texttt{Search queries, gift messages}

\texttt{Only skip pure UI labels like "Price:", "Quantity:", "Add to cart".}\\
\texttt{When in doubt, include it.}

\texttt{Return JSON: \{"pii\_items": [\{"text": "exact text"\}]\}}
\end{quote}

\subsection{Per-Class Performance Breakdown}

Table~\ref{tab:per_class_full} shows per-class AP@50 on fully-filled test images across split strategies for 2-class detection.

\begin{table}[h]
\caption{Per-class AP@50 on fully-filled test images (test\_full variant).}
\label{tab:per_class_full}
\begin{center}
\begin{small}
\begin{sc}
\begin{tabular}{lcc}
\toprule
Split & Text & Image \\
\midrule
Cross-Page & 0.623 & 0.870 \\
Cross-Company & 0.471 & 0.855 \\
Cross-Type & 0.552 & 0.875 \\
\bottomrule
\end{tabular}
\end{sc}
\end{small}
\end{center}
\vskip -0.1in
\end{table}


Product images achieve robust detection across all splits (0.855--0.875 AP@50), confirming that visual elements have distinctive signatures that generalize well regardless of company branding or page type. Text detection shows more variation: cross-page performance reaches 0.623 AP@50, while cross-company and cross-type splits achieve 0.471 and 0.552 AP@50 respectively. This degradation reflects the challenge of generalizing text field detection when visual styling (borders, fonts, spacing) varies across unseen companies or form types.

\subsection{Split Strategy Ablation}
\label{app:split_fill_ablation}

Table~\ref{tab:split_comparison} compares three split strategies that evaluate generalization at different levels. Test$_{\text{Cross-Page}}$ achieves the strongest average performance (0.797 mAP@50), with particularly high partial-fill detection (0.842), indicating that models learn layout-invariant features within a company's design system. Test$_{\text{Cross-Company}}$ degrades to 0.753 when generalizing to Amazon's distinct visual style, with the largest drop on full-fill images (0.663) where dense product layouts and unique styling present the greatest challenge. Test$_{\text{Cross-Type}}$ shows the most uniform degradation across fill states (0.713--0.744), revealing that page-type-specific patterns---such as dense tabular receipts versus interactive checkout forms---transfer less effectively than company-specific design conventions.

\begin{table}[h]
  \caption{Split strategy comparison (\textsc{WebRedact}, 2-class, trained on full+partial). Test$_{\text{Cross-Page}}$ performs best, while Test$_{\text{Cross-Type}}$ presents the strongest generalization challenge.}
  \label{tab:split_comparison}
  \begin{center}
  \begin{small}
  \begin{sc}
  \begin{tabular}{lcccc}
  \toprule
  Split Strategy & full & partial & empty & Avg \\
  \midrule
  Test$_{\text{Cross-Page}}$ & 0.747 & 0.842 & 0.803 & \textbf{0.797} \\
  Test$_{\text{Cross-Company}}$ & 0.663 & 0.839 & 0.756 & 0.753 \\
  Test$_{\text{Cross-Type}}$ & 0.713 & 0.727 & 0.744 & 0.728 \\
  \bottomrule
  \end{tabular}
  \end{sc}
  \end{small}
  \end{center}
  \vskip -0.1in
\end{table}

\subsection{Fill State Ablation}

Table~\ref{tab:fill_ablation} ablates training data composition to isolate the contribution of each fill state. Training on full screenshots alone achieves 0.771 mAP@50. Counterintuitively, adding empty screenshots \emph{without} partials degrades performance to 0.758, likely because the visual gap between empty forms (placeholder text, unfilled fields) and fully-filled forms is too large for the model to bridge without intermediate examples. Adding partial-fill data resolves this: full+partial achieves 0.797 (+0.026 over full-only), with the strongest gains on partial test images (0.842 versus 0.810), confirming that mid-entry states provide essential visual grounding. Combining all three states achieves the best overall performance (0.825 mAP@50), demonstrating that empty forms become useful once partial fills provide the intermediate signal that bridges the visual gap.

\begin{table}[h]
  \caption{Fill state ablation on Test$_{\text{Cross-Page}}$ (\textsc{WebRedact}, 2-class). Partial-fill data is essential; empty screenshots help only when combined with partials.}
  \label{tab:fill_ablation}
  \begin{center}
  \begin{small}
  \begin{sc}
  \begin{tabular}{lcccc}
  \toprule
  Train & full & part. & empty & Avg \\
  \midrule
  empty & 0.672 & 0.818 & 0.777 & 0.756 \\
  full & 0.736 & 0.810 & 0.767 & 0.771 \\
  full+empty & 0.734 & 0.774 & 0.768 & 0.758 \\
  full+part. & 0.747 & 0.842 & 0.803 & 0.797 \\
  full+part.+empty & \textbf{0.762} & \textbf{0.881} & \textbf{0.831} & \textbf{0.825} \\
  \bottomrule
  \end{tabular}
  \end{sc}
  \end{small}
  \end{center}
  \vskip -0.1in
\end{table}

\subsection{Progressive Fill Density Ablation}
\label{app:progressive_fill_ablation}

We ablate the number of partial-fill screenshots per layout to measure how progressive completion stages affect model performance. Each layout generates multiple partial-fill variants representing different stages of form completion. We evaluate 1, 3, and 5 partial stages; higher sampling counts would effectively approximate capturing all possible intermediate states (e.g., sampling 7 partials from a 7-field form would capture every field-by-field completion step).

\begin{table}[h]
\centering
\captionsetup{width=0.85\textwidth}
\caption{Progressive fill density ablation (\textsc{WebRedact}, 2-class, cross-page, trained on full+partial). Shows per-variant breakdown and overall metrics.}
\label{tab:progressive_fill_ablation}
\begin{small}
\begin{sc}
\begin{tabular}{lccccccc}
\toprule
Partials & Train & test\_full & test\_partial & test\_empty & Avg mAP & Prec. & Rec. \\
\midrule
1 & 521 & 0.727 & 0.774 & 0.774 & 0.758 & 0.793 & 0.688 \\
3 & 763 & 0.738 & 0.794 & 0.779 & 0.770 & 0.811 & 0.709 \\
5 & 938 & 0.762 & 0.835 & 0.810 & \textbf{0.802} & 0.822 & 0.727 \\
\bottomrule
\end{tabular}
\end{sc}
\end{small}
\vskip -0.1in
\end{table}

Performance improves consistently with more partial stages, increasing from 0.758 mAP@50 (1 partial) to 0.802 (5 partials), with gains across all test variants. Both precision and recall improve (0.793 to 0.822 and 0.688 to 0.727 respectively), demonstrating that each progressive completion stage provides non-redundant learning signal. The model shows particularly strong gains on test\_partial images (0.774 to 0.835), confirming that training on multiple intermediate form-filling states enables more robust field detection patterns.

\subsection{Data Variant Ablation Details}
\label{app:variant_ablation}

We ablate the number of text variants per layout to measure how data diversity affects generalization. Each unique layout is rendered with 1, 10, or 25 distinct PII data injections (varying names, addresses, products, etc.).

\begin{table}[h]
\centering
\captionsetup{width=0.85\textwidth}
\caption{Text variant ablation on Test$_{\text{Cross-Page}}$ (\textsc{WebRedact}, 2-class, trained on full+partial). Shows per-variant breakdown and overall metrics.}
\label{tab:variant_ablation_full}
\begin{small}
\begin{sc}
\begin{tabular}{lcccccc}
\toprule
Variants & test\_full & test\_partial & test\_empty & Avg mAP & Prec. & Rec. \\
\midrule
1 & 0.762 & 0.831 & 0.791 & 0.795 & 0.805 & 0.713 \\
10 & 0.774 & 0.856 & 0.804 & 0.811 & 0.829 & 0.722 \\
25 & 0.783 & 0.866 & 0.812 & \textbf{0.820} & 0.842 & 0.731 \\
\bottomrule
\end{tabular}
\end{sc}
\end{small}
\vskip -0.1in
\end{table}

Performance improves consistently with more text variants, rising through 0.795, 0.811, and 0.820, demonstrating that the injectable data system successfully provides training diversity beyond layout diversity alone. Both precision and recall increase with more variants (0.805 to 0.842 and 0.713 to 0.731 respectively). The model shows particularly strong gains on test\_partial images (0.831 to 0.866), confirming that diverse PII data injections (varying names, addresses, products) enable the model to learn more robust patterns that generalize across different data instances.

\subsection{Qualitative Prediction Analysis}

\begin{figure*}[!t]
\centering
\includegraphics[width=0.9\textwidth, trim=0 150 0 250, clip]{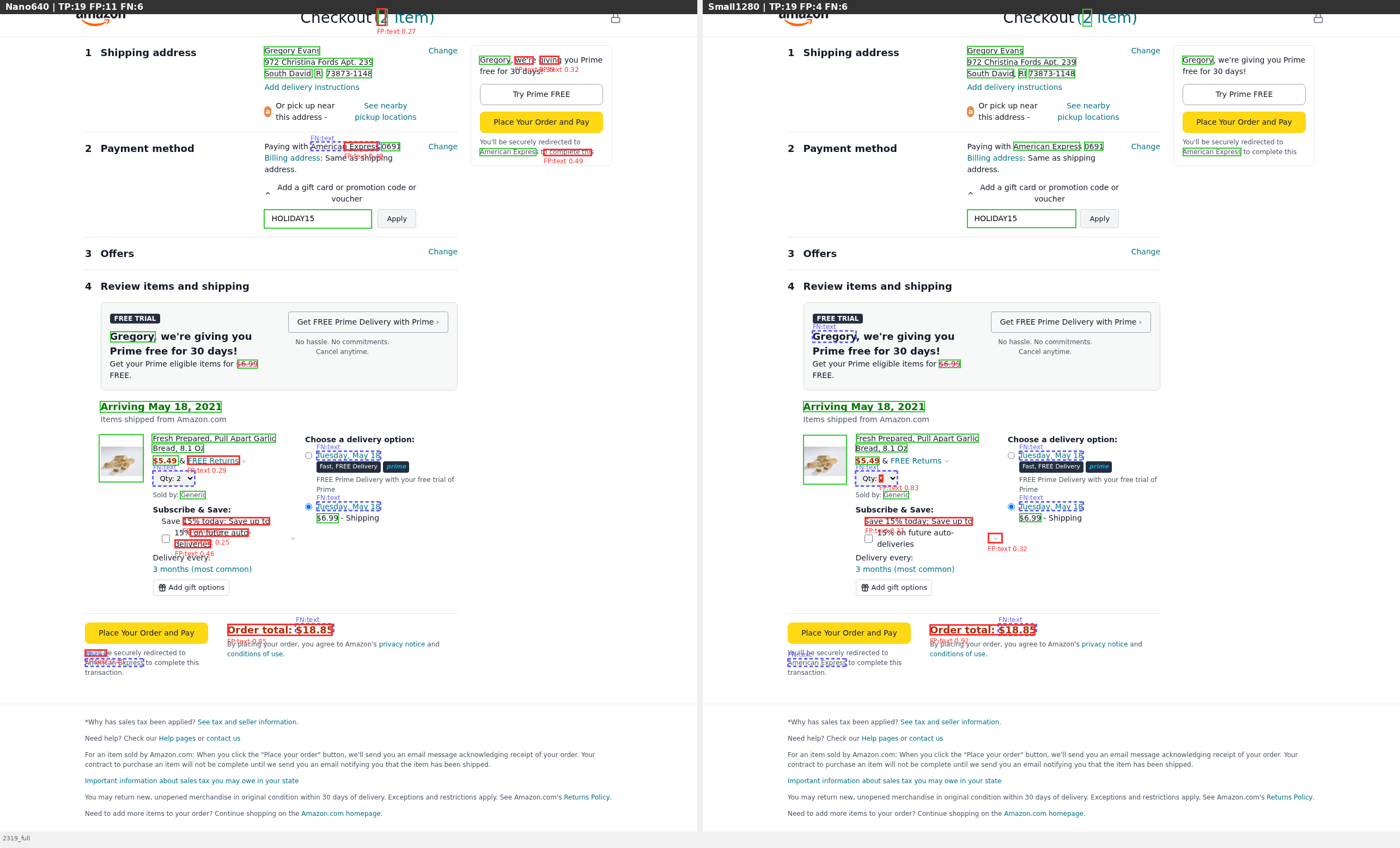}\\[1em]
\includegraphics[width=0.9\textwidth, trim=0 800 0 40, clip]{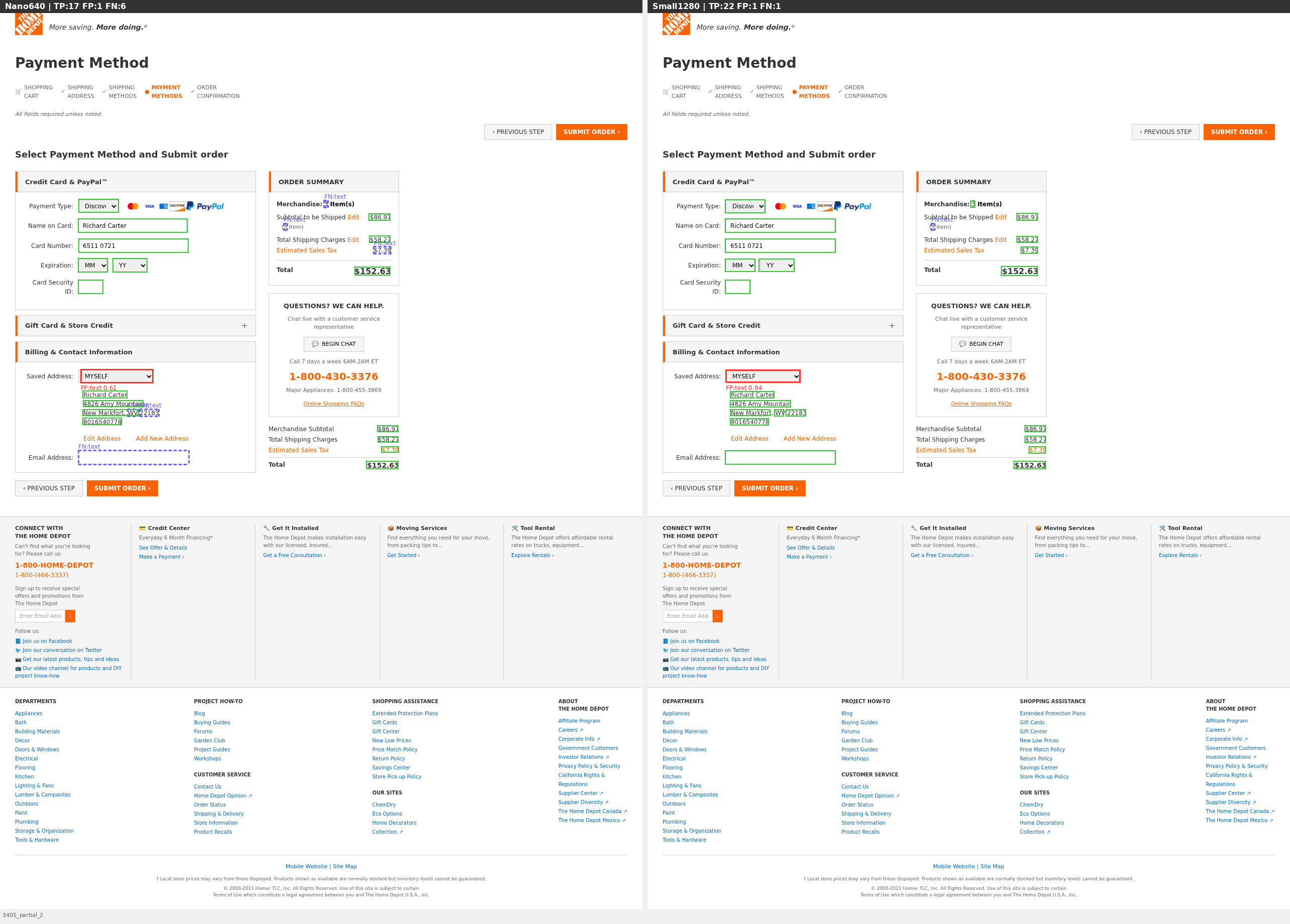}
\caption{Qualitative comparison of \textsc{WebRedact} (left) and \textsc{WebRedact-large} (right) predictions on cross-page test images. Green boxes indicate correct alignment with ground truth, red boxes indicate false positives, and blue boxes indicate false negatives.}
\label{fig:failure_modes}
\end{figure*}

Figure~\ref{fig:failure_modes} shows prediction patterns for \textsc{WebRedact} and \textsc{WebRedact-large} on cross-page test images. Both models demonstrate strong performance on core detection targets: all input fields, product images, and prices  achieve near-perfect detection. These well-represented patterns in the training data are reliably captured.

Failures predominantly occur on underrepresented edge cases. The top image exhibits numerous false positives (red boxes), where the model over-detects elements that the ground truth does not annotate. \textsc{WebRedact} also annotates promotional text like ``Save up to 15\% on future auto deliveries'', which represents marketing content rather than user-specific information. These false positives indicate the model responds to certain visual contexts (callouts, banners, emphasized text) where distinguishing promotional content from actual data fields remains challenging. Missed detections (blue boxes) occur when PII appears in atypical visual presentations: the recipient name ``Gregory'' rendered as a large bold callout at the top of the page rather than within a standard form field, and delivery dates like ``Tuesday, May 18'' styled in green text rather than conventional black typography. The bottom image shows detection patterns on a payment page, where both models successfully capture most standard form fields but exhibit characteristic failures. \textsc{WebRedact} misses hard-to-notice form fields without observable borders when empty, and both models struggle with small numbers in densely packed locations like ``2 item'' quantity indicators. Both models also produce false positives on an unlabeled dropdown menu—though this reflects legitimate annotation ambiguity, as the same dropdown could reasonably be tagged if rendered with a visible selected value. Despite these edge case failures, form field detection remains robust overall. These failures reflect the dataset's composition: while training data spans diverse e-commerce contexts, forms and cart pages dominate and establish the primary visual patterns for input fields, product images, and prices. The model learns these dominant patterns robustly but struggles when sensitive information appears in more information-dense or ambiguous contexts.

To assess generalization to non-synthetic data, we qualitatively evaluated the model on the original e-commerce screenshots used as generation targets, as well as 50 additional real e-commerce screenshots that were never synthetically reproduced from the same brands. We do not release these images due to their sensitive content. Upon observing the annotations, we recognize that input fields, prices, and product details demonstrate robust detection across varied contexts, confirming that patterns learned from synthetic reproductions transfer to authentic interfaces. However, shipping dates, cardholder names rendered on card images, and basket quantities exhibit inconsistent detection, consistent with the underrepresentation patterns observed on synthetic test data demonstrated in Figure~\ref{fig:failure_modes}.

\end{document}